\documentclass[a4paper,onecolumn,11pt,accepted=2024-10-15]{quantumarticle}

\pdfoutput=1 

\usepackage{amsmath}
\usepackage{amsfonts}
\usepackage[margin=1in]{geometry}
\usepackage{graphicx}
\usepackage{hyperref}
\usepackage{url} 
\usepackage{bm}
\usepackage{tikz}
\usepackage[subrefformat=parens]{subcaption}
\usepackage{ifthen} 
\usepackage[T1]{fontenc}
\usepackage[title]{appendix} 
\usepackage{thmtools,thm-restate} 
\usepackage{doi} 
\usepackage{cuted} 
\usepackage{gensymb} 

\usepackage{amsthm} 
\usepackage{mdframed} 

\makeatletter
\def\thmhead@plain#1#2#3{%
	\thmname{#1}\thmnumber{\@ifnotempty{#1}{ }\@upn{#2}}%
	\thmnote{ {\the\thm@notefont#3}}}
\let\thmhead\thmhead@plain
\def\th@definition{%
	\thm@notefont{}
	\normalfont 
}
\makeatother

\theoremstyle{definition}
\newmdtheoremenv{definition}{Definition}

\newcommand{\li}{\bar{\mathbb I}}
\newcommand{\lx}{\bar{X}}
\newcommand{\ly}{\bar{Y}}
\newcommand{\lz}{\bar{Z}}

\newcommand{\pr}[1]{\operatorname{Pr}\left(#1\right)}

\newcommand{\ie}{i.e., }
\newcommand{\eg}{e.g., }
\newcommand{\orr}{\mathrm{OR}}
\newcommand{\andd}{\mathrm{AND}}
\newcommand{\nott}{\mathrm{NOT}}
\newcommand{\true}{\mathrm{TRUE}}
\newcommand{\false}{\mathrm{FALSE}}
\newcommand{\p}{\mathrm{P}}
\newcommand{\np}{\mathrm{NP}}
\newcommand{\fp}{\mathrm{FP}}
\newcommand{\sharpP}{\mathrm{\#P}}

\newcommand{\defbox}[1]{
	\noindent\fbox{\hspace{.03\linewidth}
		\begin{minipage}{.9\linewidth}
			\vspace{.03\linewidth} #1 \vspace{.01\linewidth}
		\end{minipage}\hspace{.03\linewidth}
	}
	\vspace{11pt}
}

\definecolor{zStabColor}{rgb}{0.7, 0.7, 0.9} 
\definecolor{xStabColor}{rgb}{0.8, 1.0, 0.8} 
\definecolor{noiseColor}{rgb}{0.8, 0, 0.8} 
\definecolor{errorColor}{rgb}{1.0, 0, 0} 
\definecolor{locationLabelColor}{rgb}{1.0, 0.3, 0} 
\definecolor{labelColor}{rgb}{0, 0, 0} 
\definecolor{syndromeColor}{rgb}{0, 0, 0} 

\tikzset{
	noiseStyle/.style={text=noiseColor,font=\scriptsize\sffamily}
}

\tikzset{
	errorStyle/.style={text=errorColor,font=\normalsize\sffamily, node font=\bfseries}
}

\tikzset{
	locationLabelStyle/.style={text=locationLabelColor,font=\small}
}

\tikzset{
	labelStyle/.style={align=center,text=labelColor,font=\small}
}

\tikzset{
	syndromeStyle/.style={text=syndromeColor,font=\scriptsize}
}

\tikzset{
	qubitStyle/.style={circle,fill=black, inner sep=2pt}
}

\newcommand{\surfaceCodeCheckerboard}[5]{
	\foreach \x in {2,...,#1} {
		\foreach \y in {2,...,#2} {
			\pgfmathsetmacro\stabColor{iseven(\x+\y) ? "zStabColor" : "xStabColor"}
			\pgfmathsetmacro\lineColor{#3 ? "black" : "\stabColor"}
			\draw[fill=\stabColor, draw=\lineColor] (\x-2, \y-2) rectangle ++(1,1);
		}
	}
	
	\pgfmathsetmacro{\bool}{#5 ? 1 : 0};
	\ifthenelse{\bool=1}{
		\pgfmathsetmacro\lineColor{#3 ? "black" : "xStabColor"}
		
		\pgfmathsetmacro{\endLoop}{#1-2};
		\foreach \x in {0,2,...,\endLoop} {
			\filldraw[draw=\lineColor,fill=xStabColor]
				(\x, 0) arc (-180:0:0.5) -- cycle;
		}
		
		\pgfmathsetmacro{\startLoop}{iseven(#2) ? 0 : 1};
		\pgfmathsetmacro{\startLoopPlusTwo}{\startLoop+2};
		\foreach \x in {\startLoop,\startLoopPlusTwo,...,\endLoop} {
			\filldraw[draw=\lineColor,fill=xStabColor]
			(\x, #2-1) arc (180:0:0.5) -- cycle; 
		}
		
		\pgfmathsetmacro\lineColor{#3 ? "black" : "zStabColor"}
		
		\pgfmathsetmacro{\endLoop}{#2-2};
		\foreach \y in {1,3,...,\endLoop} {
			\filldraw[draw=\lineColor,fill=zStabColor]
			(0, \y) arc (90:-90:-0.5) -- cycle;
		}
		\pgfmathsetmacro{\startLoop}{iseven(#1) ? 1 : 0};
		\pgfmathsetmacro{\startLoopPlusTwo}{\startLoop+2};
		\foreach \y in {\startLoop,\startLoopPlusTwo,...,\endLoop} {
			\filldraw[draw=\lineColor,fill=zStabColor]
			(#1-1, \y) arc (-90:90:0.5) -- cycle;
		}
	}{}
	
	\pgfmathsetmacro{\bool}{#4 ? 1 : 0};
	\ifthenelse{\bool=1}{
		\foreach \x in {1,...,#1} {
			\foreach \y in {1,...,#2} {
				\node at (\x-1,\y-1) [qubitStyle] {};
			}
		}
	}{}
}

\newcommand{\andGadget}{
\surfaceCodeCheckerboard{30}{27}{false}{false}{false}

\node[syndromeStyle] at (12.5, 19.5) {$-1$};

\node at (12,21) [noiseStyle] {Y};

\foreach \y in {22,23} \node at (12,\y) [noiseStyle] {Z};
\draw[draw=black,line width=1pt] (10,23.3) rectangle ++(4,1.35);
\node [labelStyle,font=\scriptsize] at (12, 24) {convert to \\ $X$ string};
\foreach \y in {25,26} \node at (12,\y) [noiseStyle] {X};

\node [labelStyle] at (12, 26.5) {output wire};

\node [labelStyle] at (3, -0.5) {input 1};
\foreach \y in {0,1,2}  \node at (3, \y) [noiseStyle] {X};
\node at (3, 3) [noiseStyle] {X,Z};
\node at (3, 4) [noiseStyle] {X};
\node at (3, 5) [noiseStyle] {X,Y};
\node at (3, 6) [noiseStyle] {X};
\node at (3, 7) [noiseStyle] {X,Y};
\foreach \y in {8,...,11} \node at (3, \y) [noiseStyle] {X};
\node at (3, 12) [noiseStyle] {X,Z};
\foreach \y in {13,...,21} \node at (3, \y) [noiseStyle] {X};
\foreach \x in {4,...,10} \node at (\x, 21) [noiseStyle] {X};
\node at (11,21) [noiseStyle] {X,Z};

\draw[draw=black,line width=1pt] (6.5, 11.5) rectangle ++(5, 6);
\node [labelStyle] at (7, 18.4) {this is a convert to \\ $Z$ string gadget};

\foreach \x in {4,...,9} \node at (\x, 9) [noiseStyle] {X};
\foreach \y in {10,11,12} \node at (10,\y) [noiseStyle] {X};
\node at (10,13) [noiseStyle] {Y};
\foreach \y in {14,15} \node at (10,\y) [noiseStyle] {Z};
\node at (10,16) [noiseStyle] {Y};
\node at (10,17) [noiseStyle] {Z};
\node at (11,18) [noiseStyle] {Z};
\node at (12,19) [noiseStyle] {Z};

\foreach \x in {9,8} \node at (\x,13) [noiseStyle] {X};
\foreach \y in {14,15} \node at (7,\y) [noiseStyle] {X};
\foreach \x in {8,9} \node at (\x,16) [noiseStyle] {X};
\node at (11,16) [noiseStyle] {X};
\node at (12,15) [noiseStyle] {X};
\foreach \y in {14,...,8} \node at (12,\y) [noiseStyle] {X};
\node at (12,6) [noiseStyle] {X};
\node at (12,4) [noiseStyle] {X};
\node at (12,3) [noiseStyle] {X,Z};
\foreach \y in {2,1} \node at (12,\y) [noiseStyle] {X};
\node [labelStyle] at (8.4, 0.5) {to bottom boundary $\downarrow$};

\foreach \x in {9,...,4} \node at (\x, 12) [noiseStyle] {Z};
\foreach \x in {1,2} \node at (\x,12) [noiseStyle] {Z};
\foreach \y in {11,...,7} \node at (1,\y) [noiseStyle] {Z};
\node at (2,7) [noiseStyle] {Z};
\foreach \x in {4,...,11} \node at (\x,7) [noiseStyle] {Z};
\node at (12,7) [noiseStyle] {Y};
\node at (13,7) [noiseStyle] {Z};
\node at (14,7) [noiseStyle] {X,Z};
\foreach \x in {15,...,23} \node at (\x,7) [noiseStyle] {Z};
\node at (25, 7) [noiseStyle] {Z};
\node at (26, 6) [noiseStyle] {Z};
\foreach \x in {26,25} \node at (\x,5) [noiseStyle] {Z};
\foreach \x in {23,...,15} \node at (\x,5) [noiseStyle] {Z};
\node at (14,5) [noiseStyle] {X,Z};
\node at (13,5) [noiseStyle] {Z};
\node at (12,5) [noiseStyle] {Y};
\foreach \x in {11,...,4} \node at (\x,5) [noiseStyle] {Z};
\foreach \x in {2,1} \node at (\x,5) [noiseStyle] {Z};
\node [labelStyle] at (-0.2, 5.5) {to left boundary \\ $\longleftarrow$};

\node at (12,20) [noiseStyle] {Z};
\node at (10,22) [noiseStyle] {Z};
\node at (9,23) [noiseStyle] {Z};
\foreach \x in {8,7,6} \node at (\x,23) [noiseStyle] {Z};
\node [labelStyle] at (2.2, 23) {to left boundary $\longleftarrow$};

\node [labelStyle] at (24, -0.5) {input 2};
\foreach \y in {0,1,2}  \node at (24, \y) [noiseStyle] {X};
\node at (24, 3) [noiseStyle] {X,Y};
\node at (24, 4) [noiseStyle] {X};
\node at (24, 5) [noiseStyle] {X,Z};
\node at (24, 6) [noiseStyle] {X};
\node at (24, 7) [noiseStyle] {X,Z};
\foreach \y in {8,...,12} \node at (24, \y) [noiseStyle] {X};
\node at (24, 13) [noiseStyle] {X,Z};
\foreach \y in {14,...,21} \node at (24, \y) [noiseStyle] {X};
\foreach \x in {23,...,13} \node at (\x, 21) [noiseStyle] {X};

\draw[draw=black,line width=1pt] (14.5, 12.5) rectangle ++(5, 6);
\node [labelStyle] at (20, 19.3) {this is a convert to \\ $Z$ string gadget};

\foreach \x in {23,...,16} \node at (\x, 9) [noiseStyle] {X};
\foreach \y in {10,...,13} \node at (16, \y) [noiseStyle] {X};
\node at (16, 14) [noiseStyle] {Y};
\foreach \y in {15,16} \node at (16, \y) [noiseStyle] {Z};
\node at (16, 17) [noiseStyle] {Y};
\foreach \y in {18,19,20} \node at (16, \y) [noiseStyle] {Z};
\foreach \x in {15,14,13} \node at (\x, 20) [noiseStyle] {Z};

\foreach \x in {17,...,23} \node at (\x, 13) [noiseStyle] {Z};
\foreach \x in {25,26,27} \node at (\x, 13) [noiseStyle] {Z};
\foreach \y in {12,...,3} \node at (28, \y) [noiseStyle] {Z};
\foreach \x in {27,26,25} \node at (\x, 3) [noiseStyle] {Z};
\foreach \x in {23,...,15} \node at (\x, 3) [noiseStyle] {Z};
\node at (14, 3) [noiseStyle] {Y};
\node at (13, 3) [noiseStyle] {Z};
\foreach \x in {11,...,4} \node at (\x, 3) [noiseStyle] {Z};
\foreach \x in {2,1} \node at (\x, 3) [noiseStyle] {Z};
\node [labelStyle] at (-0.2, 3.5) {to left boundary \\ $\longleftarrow$};

\foreach \x in {17,18} \node at (\x, 14) [noiseStyle] {X};
\foreach \y in {15,16} \node at (19,\y) [noiseStyle] {X};
\foreach \x in {18,17} \node at (\x, 17) [noiseStyle] {X};
\node at (15, 17) [noiseStyle] {X};
\foreach \y in {16,...,8} \node at (14,\y) [noiseStyle] {X};
\node at (14,6) [noiseStyle] {X};
\node at (14,4) [noiseStyle] {X};
\foreach \y in {2,1} \node at (14,\y) [noiseStyle] {X};

\node [labelStyle] at (17.6, 0.5) {$\downarrow$ to bottom boundary};
}
\newcommand{\figref}[1]{Fig.~\ref{#1}}
\newcommand{\appref}[1]{Appendix \ref{#1}}
\newcommand{\defref}[1]{definition \ref{#1}}

\begin{document}
	
\title{Hardness results for decoding the surface code with Pauli noise}
\author{Alex Fischer}
\email{alexander.fischer3@gmail.com}
\author{Akimasa Miyake}
\email{amiyake@unm.edu}
\affiliation{Department of Physics and Astronomy, Center for Quantum Information and Control,
	University of New Mexico, Albuquerque, New Mexico 87106, USA}
\maketitle

\begin{abstract}
	Real quantum computers will be subject to complicated, qubit-dependent noise, instead of simple noise such as depolarizing noise with the same strength for all qubits. We can do quantum error correction more effectively if our decoding algorithms take into account this prior information about the specific noise present. This motivates us to consider the complexity of surface code decoding where the input to the decoding problem is not only the syndrome-measurement results, but also a noise model in the form of probabilities of single-qubit Pauli errors for every qubit.
	
	In this setting, we show that quantum maximum likelihood decoding (QMLD) and degenerate quantum maximum likelihood decoding (DQMLD) for the surface code are NP-hard and \#P-hard, respectively. We reduce directly from SAT for QMLD, and from \#SAT for DQMLD, by showing how to transform a boolean formula into a qubit-dependent Pauli noise model and set of syndromes that encode the satisfiability properties of the formula. We also give hardness of approximation results for QMLD and DQMLD. These are worst-case hardness results that do not contradict the empirical fact that many efficient surface code decoders are correct in the average case (\ie for most sets of syndromes and for most reasonable noise models). These hardness results are nicely analogous with the known hardness results for QMLD and DQMLD for arbitrary stabilizer codes with independent $X$ and $Z$ noise.
\end{abstract}

\tableofcontents

\section{Introduction}
	Quantum computers need to perform long, error-free computations in order to solve problems of interest. All the physical qubits we can build now (and all those we are likely to be able to build in the future) have error rates that are much too high to run long computations error-free. Therefore in order to run computations of interest with these physical qubits, we have to use quantum error correction to do fault-tolerant quantum computation. This involves encoding logical qubits with many physical qubits using quantum error correcting codes, performing computations by using logical operations that act on the logical qubits, and continuously correcting errors on the encoded logical qubits as they occur throughout the computation.
	
	The surface code\cite{surfaceCodeBoundaryBravyiKitaev} is one of the most promising candidates for a quantum error correcting code on which to do large-scale fault-tolerant quantum computation, due to its high error thresholds and locality of all operations necessary for fault tolerance. The locality of the error correction operations is important because many physical implementations of qubits, such as superconducting qubits and other solid-state qubits, directly support only local, nearest-neighbor 2-qubit operations. Fault-tolerant quantum computers made with solid-state qubits are likely to use surface codes or related topological codes\cite{ogorman2014surfaceCodeSilicon,hill2015surfaceCodeSilicon,gambetta2015surfaceCodeSuperconducting}. Small versions of the surface code have been experimentally implemented on several superconducting quantum computers\cite{firstSurfaceCodeExperiment21,ZuchongzhiSurfaceCodeExperiment21,Google2022suppressing}.
	
	Decoding is the computational task of determining what error correction operations to apply, given a set of syndrome measurement results extracted from the code. For superconducting qubits, syndromes are measured (and need to be decoded) on the order of once every microsecond\cite{firstSurfaceCodeExperiment21,ZuchongzhiSurfaceCodeExperiment21,Google2022suppressing}. Additionally, it may be required to use surface codes with up to thousands\cite{gidney2021factor} or even tens of thousands\cite{surfaceCodeReviewFowler} of physical qubits per logical qubit in order to perform quantum computations of interest. Processing thousands of syndromes per microsecond (per logical qubit) is a formidable computational task that has motivated the study of efficient and accurate decoding algorithms for the surface code, and for quantum codes in general.
	
	There are very few special cases where any surface code decoding algorithms are known to be optimal or even approximately optimal in any rigorous sense. Existing decoding algorithms either make simplifying assumptions about the noise (such as independence of $X$ and $Z$ noise\cite{topologicalQuantumMemory,BSV14}, which is a particularly useful assumption for CSS codes such as the surface code because CSS codes treat $X$ and $Z$ errors separately) that allow them to decode optimally, or are heuristic algorithms without rigorous performance guarantees. See \cite{surfaceCodeDecodersReview23} for a review of surface code decoding algorithms. Given all the effort put into efficient decoders for the surface code, the lack of success in finding provably optimal algorithms suggests that there may be computational complexity reasons blocking progress. In this paper we establish computational complexity results that explain this lack of progress.
	
	Related work on the complexity of non-degenerate and degenerate quantum maximum likelihood decoding has focused on general stabilizer codes, rather than the surface code. Non-degenerate quantum maximum likelihood decoding, or just quantum maximum likelihood decoding (QMLD), is the task of finding the most likely error to have occurred, given the syndrome measurement results. Degenerate quantum maximum likelihood decoding (DQMLD) is the task of finding a correction operation that maximizes the probability that we correct the error, taking into account that multiple different errors can be corrected by the same correction operation, an important property of quantum codes called degeneracy. DQMLD is the superior decoding strategy because it maximizes the probability that we correct the error, which is what we really care about. However, DQMLD is generally more computationally intensive than QMLD, so QMLD may be used in practice. Degenerate quantum maximum likelihood decoding (DQMLD) is sometimes just called maximum likelihood decoding, especially in the context of the surface code (\eg in \cite{BSV14}). See Section \ref{QMLDvsDQMLD} for precise definitions of QMLD and DQMLD.
	
	Decoding general stabilizer codes is known to be NP-hard for QMLD and \#P-hard for DQMLD. Hsieh and Le Gall\cite{npHardnessDecodingQuantumHsiehLeGall11} showed that both QMLD and DQMLD for general stabilizer codes are NP-hard when every qubit has independent $X$ and $Z$ noise with some constant probability, by reducing from the problem of decoding classical binary linear codes, which is also NP-hard\cite{classicalDecodingHard78}. Kuo and Lu\cite{hardnessDecodingDepolarizingKuoLu12,hardnessDecodingDepolarizingKuoLu20} showed that QMLD and DQMLD for stabilizer codes remain NP-hard with depolarizing noise. Iyer and Poulin\cite{hardnessStabilizerCodesIyerPoulin15} strengthened the former result by showing that DQMLD for stabilizer codes with independent $X$ and $Z$ noise is actually \#P-hard, not merely NP-hard. \#P is the complexity class of function problems that count the number of accepting paths for a nondeterministic Turing machine. The canonical complete problem for \#P is \#SAT: counting the number of satisfying assignments of a boolean formula.
	
	We emphasize that our results are neither a stronger nor weaker version of these prior results; our results are about related but distinct decoding problems. Prior work considers the complexity of decoding where the input to the computational problem is the stabilizer generators of the code being considered and the measurement results for those stabilizer generators (\ie syndromes), while the noise model is fixed to either independent $X$ and $Z$ noise or depolarizing noise. Our work considers the complexity of decoding where the input to the computational problem is the syndromes and a description of the noise model in the form of probabilities of Pauli errors for each qubit, while the stabilizers are fixed to those of the surface code.
	
	These prior results show that decoding general stabilizer codes is hard by embedding a hard computational problem into a set of stabilizer generators and measurement results for those stabilizer generators. These prior results show that an optimal decoding algorithm that works for any quantum stabilizer code, but with some fixed noise model (either independent $X$ and $Z$ noise, or depolarizing noise) can be used to solve NP-hard or \#P-hard problems. Our results on the hardness of decoding the surface code instead show that an optimal decoding algorithm that only works on the surface code, but importantly, works for any Pauli noise model, can be used to solve NP-hard or \#P-hard problems. We need the Pauli noise model ingredient for our results; because we don't have the freedom to choose stabilizers in a way that cleverly encodes a hard computational problem, we instead choose a Pauli noise model that encodes a hard computational problem in our reductions.
	
	Much of the research on decoders for the surface code and related topological codes has focused on algorithms that use information about the specific noise present in order to decode better (see \cite{surfaceCodeDecodersReview23} for examples). Real quantum computers will not be subject to simple noise such as depolarizing noise, but rather more complicated noise that may be different for each qubit. We would like our decoding algorithms to take advantage of this fact, and in fact there are many decoding algorithms that take as input not only the syndromes but also a noise model, often in the form of Pauli error probabilities for each qubit. With physically realistic variation in the noise, these decoders can have much better error correction properties than what can be attained by approximating the noise with something like independent identically distributed depolarizing noise\cite{inidNoiseSurfaceCode,ninidNoiseSurfaceCode}, so these decoders will play an important role in building large fault-tolerant quantum computers. Therefore it is natural to consider the hardness of surface code decoding where the input to the problem is not only the syndromes but also Pauli error probabilities for each qubit. This is the exact problem we show to be hard in this work.
	
	Decoding the surface code is known to be closely related computing partition functions of Ising models, which is known to be \#P-hard\cite{JerrumSinclair93IsingApproximation}, following Barahona's seminal result on its NP-hardness\cite{barahonaIsingComplexity}. Specifically, the coset probability of an error is exactly the partition function of an Ising model constructed from that error and the noise model\cite{topologicalQuantumMemory,topologicalCodesDepolarizationResilience,ChubbFlammiaStatMechSurfaceCode}. So computing partition functions allows one to determine which error coset has maximum coset probability, \ie to solve DQMLD for the surface code. However, this does not directly imply any hardness results for DQMLD for the surface code, since the task for DQMLD is merely to decide which error coset has the greatest coset probability (\ie which partition function out of several related ones is the greatest), rather than to exactly compute any coset probabilities.
	\subsection{Summary of results}
		We consider the complexity of surface code decoding where every qubit $j$ has a Pauli $X$ error with probability $p_X^{(j)}$, a Pauli $Y$ error with probability $p_Y^{(j)}$, a Pauli $Z$ error with probability $p_Z^{(j)}$, and no error with probability $1-p_X^{(j)}-p_Y^{(j)}-p_Z^{(j)}$. In general the probabilities $p_X^{(j)}$, $p_Y^{(j)}$, $p_Z^{(j)}$ will be different for each qubit $j$, and they are part of the input to the computational problem along with the syndromes. Errors for different qubits are independent. We prove 2 theorems about the worst-case hardness of decoding the surface code with this simple but general class of noise models, called independent non-identically distributed Pauli noise, or i.n.i.d.~Pauli noise (introduced in \cite{inidNoiseSurfaceCode}).
		\begin{restatable*}{theorem}{qmldHardTheorem}
			\label{qmldHardTheorem}
			Quantum maximum likelihood decoding (QMLD) for the surface code with independent non-identically distributed Pauli noise is NP-hard.
		\end{restatable*}
		\begin{restatable*}{theorem}{dqmldHardTheorem}
			\label{dqmldHardTheorem}
			Degenerate quantum maximum likelihood decoding (DQMLD) for the surface code with independent non-identically distributed Pauli noise is \#P-hard under Turing reductions.
		\end{restatable*}
	
		See Section \ref{QMLDvsDQMLD} for precise definitions of QMLD and DQMLD. In Appendix \ref{completenessAppendix}, we show that these hardness results become completeness results if the decoding problems are reformulated as yes/no decision problems, instead of their more natural formulation as function problems.
	
		We prove the Theorem \ref{qmldHardTheorem} by reducing directly from SAT. The strategy of the reduction is to construct a noise model that simulates a boolean formula in the sense that there is one possible error for each possible setting of the input variables for the boolean formula. We set the error probabilities such that errors corresponding to satisfying assignments of the formula have higher probabilities than errors corresponding to unsatisfying assignments of the formula, so a QMLD algorithm will find an error corresponding to a satisfying assignment of the formula, if one exists.
		
		We prove the Theorem \ref{dqmldHardTheorem} by reducing directly from \#SAT. The noise model outputted by this reduction is similar to, but not the same as, the noise model outputted by the reduction for Theorem \ref{qmldHardTheorem}---the possible errors are the same, but the probabilities of those errors are different. We set the error probabilities such that errors corresponding to unsatisfying assignments (which are all in one error coset) all have the same probability, and likewise with errors corresponding to satisfying assignments (which are all in a different error coset). Since all these probabilities are the same, finding the error coset with maximum sum of probabilities is equivalent to counting satisfying versus unsatisfying assignments of the formula.
		
		We also prove 2 hardness of approximation results, which follow from straightforward modifications of the proofs of Theorems \ref{qmldHardTheorem} and \ref{dqmldHardTheorem}. Here the approximation measures are the probability of the error found (for Corollary \ref{qmldApproxHardCorollary}) and the coset probability of the coset found (for Corollary \ref{dqmldApproxHardCorollary}).
		\begin{restatable*}{corollary}{qmldApproxHardCorollary}
			\label{qmldApproxHardCorollary}
			Approximate QMLD of the surface code with independent non-identically distributed Pauli noise, up to any exponential approximation factor (\ie  with approximation factor $2^{\ell^c}$ for any constant $c$, where $\ell=wh$ is the number of qubits in the surface code instance), is NP-hard.
		\end{restatable*}
		\begin{restatable*}{corollary}{dqmldApproxHardCorollary}
			\label{dqmldApproxHardCorollary}
			There exists a constant $c$ such that approximate DQMLD of the surface code with independent non-identically distributed Pauli noise, with approximation factor $M(\ell)=2^{\ell^c}$, is NP-hard, where $\ell=wh$ is the number of qubits in the surface code instance.
		\end{restatable*}
	
		This means that although many decoders typically approximate the problem they are trying to solve very well, they cannot hope to approximate the problem in all scenarios (\ie for all noise models and for all sets of syndromes). If the approximation measures are $\log$ probabilities (which is a common metric that decoders work with), instead of just probabilities, then these results translate into hardness of approximation results for polynomial additive approximation factors, not exponential multiplicative approximation factors.
		
		In Section \ref{regNoiseSection}, we show that we maintain the hardness results of Theorem \ref{qmldHardTheorem} and Corollary \ref{qmldApproxHardCorollary} if we restrict all the error probabilities to be either 0, or some constant $p$, which we can choose to be any constant in $(0, 0.25]$. We also show that we maintain the hardness result of Theorem \ref{dqmldHardTheorem} if we restrict the error probabilities to be either 0, $\frac{1}{2}$, $\frac{1}{3}$, or $\frac{1}{4}$, although we lose the hardness of approximation result Corollary \ref{dqmldApproxHardCorollary} in this case.
		
		All of these hardness results are about worst-case hardness, not any type of average-case hardness. Therefore they have nothing to say about the fact that many surface code decoding algorithms tend to work well in practice; in particular, well enough to achieve fault tolerance. In order to achieve fault tolerance it is only required that decoders be correct with high probability for the problem instances that are likely to occur in practice; specifically, it is required that the probability that the decoding algorithm returns a correction operation that creates a logical error approaches zero as the code size increases. It is not required that decoders always return the right answer to the QMLD or DQMLD problem. Therefore, like all hardness results for practical problems of interest, the consequence of these results is merely to inform our search for surface code decoding algorithms: we need to search for heuristic algorithms that are often but not always correct, or search for special cases that we can solve exactly.
		
		In the main text we state and prove all of our theorems with respect to the rotated surface code. In Appendix \ref{unrotatedAppendix}, we show that all of our results also hold for the standard, unrotated surface code.
\section{Background}
	\subsection{The surface code}
		Here we review the surface code concepts necessary to understand our reductions. We do not provide a comprehensive introduction to the surface code; see \cite{surfaceCodeReviewFowler} for an excellent overview. In the main text we work exclusively with the rotated surface code variant\cite{bombin2007optimal}, because it makes some parts of our proofs easier to visualize. In Appendix \ref{unrotatedAppendix} we show that the same results hold for the standard, unrotated surface code.
		
\begin{figure*}
			\begin{subfigure}{0.5\textwidth}
				\begin{center}
					\begin{tikzpicture}[x=0.5cm,y=0.5cm]
						\surfaceCodeCheckerboard{5}{5}{true}{true}{true}
						
						\draw[fill=xStabColor, draw=black] (5.5, 0.5) rectangle ++(1,1);
						\foreach \x in {5.5, 6.5} \foreach \y in {.5, 1.5} {
							\node at (\x,\y) [qubitStyle] {};
						}
						\node at (9, 1) [labelStyle] {$= X$ stabilizer};

						\draw[fill=zStabColor, draw=black] (5.5, 2.5) rectangle ++(1,1);
						\foreach \x in {5.5, 6.5} \foreach \y in {2.5, 3.5} {
							\node at (\x,\y) [qubitStyle] {};
						}
						\node at (9, 3) [labelStyle] {$= Z$ stabilizer};
					\end{tikzpicture}
				\end{center}
				\caption{}\label{surfaceCodeDef}
			\end{subfigure} %
			\begin{subfigure}{0.5\textwidth}
				\begin{center}
					\begin{tikzpicture}[x=0.5cm,y=0.5cm]
						\surfaceCodeCheckerboard{5}{5}{false}{false}{true}
						\node at (3.5, 3.5) [syndromeStyle] {$-1$};
						\node at (1.5, 2.5) [syndromeStyle] {$-1$};
						
						\foreach \y in {0,...,3}
						\node [errorStyle] at (3, \y) {X};
						
						\foreach \x in {0,1}
						\node [errorStyle] at (\x, 3) {Z};
					\end{tikzpicture}
				\end{center}
				\caption{}\label{boundaryConditionStrings}
			\end{subfigure}
			\caption{\subref{surfaceCodeDef} The surface code with the boundary conditions we use. Qubits are at the intersections of lines. The blue (darker) faces are $Z$ stabilizers and the green (lighter) faces are $X$ stabilizers. The boundary consists of 2-qubit stabilizers: $X$ stabilizers on the bottom and top boundaries, and $Z$ stabilizers on the left and right boundaries. \subref{boundaryConditionStrings} This choice of boundary conditions has an important consequence. Strings of $X$ errors can run to the top or bottom boundary without having a $-1$ syndrome at the end of the string. Likewise, strings of $Z$ errors can run to the left or right boundary without having a $-1$ syndrome at the end of the string. Red labels are a possible error consistent with these syndromes.}
		\end{figure*}
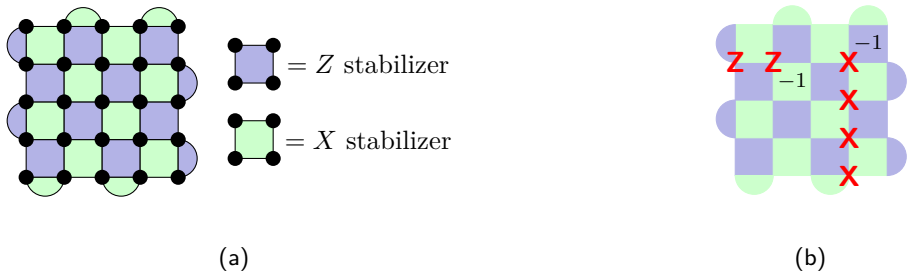

		The surface code is defined on a rectangular grid of qubits, with qubits at the intersections of grid lines. On a $w\times h$ grid of qubits, it has $n=wh$ physical qubits, $k=1$ logical qubit, and distance $d=\min(w,h)$. For every face of the grid adjacent to 4 qubits there is a 4-qubit stabilizer, either a $X^{\otimes 4}$ stabilizer or a $Z^{\otimes 4}$ stabilizer---see \figref{surfaceCodeDef}. In that figure, and all figures in this paper, green (lighter) faces indicate $X$-stabilizers and blue (darker) faces indicate $Z$ stabilizers\footnote{Readers with trouble viewing our color scheme should view our paper in black and white.}. After \figref{surfaceCodeDef}, we don't draw the lines and points on the grid, to make the figures less cluttered.
		
		The code space for the surface code is the simultaneous $+1$ eigenspace of all of those stabilizer operators. Therefore, if no error has occurred on any qubits, then all stabilizer measurement results, or \emph{syndromes}, will be $+1$. A single Pauli error $X$, $Y$, or $Z$ that occurs on a qubit causes all adjacent stabilizers that anticommute with that error to have measurement result $-1$. For example, if an $X$ ($Z$) error occurs on a qubit, then all adjacent $Z$ ($X$) stabilizers will have their measurement result flipped to $-1$. If a $Y$ error occurs on a qubit, then all adjacent stabilizers will have their measurement result flipped to $-1$.
		
		If errors occur on multiple qubits adjacent to some stabilizer, then the $\pm 1$ measurement result depends on the parity of the number of anticommuting errors adjacent to that stabilizer. If an odd number of qubits adjacent to a stabilizer have a Pauli error that anticommutes with that stabilizer, then that stabilizer will have the measurement result $-1$. If an even number of qubits adjacent to a stabilizer have a Pauli error that anticommutes with that stabilizer, then that stabilizer will have the measurement result $+1$. See \figref{boundaryConditionStrings} for an example of the stabilizer measurement results caused by a particular Pauli error.
		
		Although several choices of boundary stabilizers are possible for the surface code, here we work with the rotated surface code, which has 2-qubit $Z$ stabilizers at the left and right boundaries and 2-qubit $X$ stabilizers at the top and bottom boundaries, as in \figref{surfaceCodeDef}. This choice of boundary conditions results in a code that encodes $k=1$ logical qubit. These boundary conditions have another important consequence: strings of $X$ ($Z$) errors can run to the top or bottom (left or right) boundary without having a $-1$ syndrome at that boundary endpoint. See \figref{boundaryConditionStrings} for an example. Another important consequence is that strings of $X$ ($Z$) errors that start and end at the same top/bottom (left/right) boundary are products of stabilizers. This contrasts with the toric code with its periodic boundary conditions; the toric encodes $k=2$ logical qubits, and products of its stabilizers form closed loops of errors.
		
		The standard choice of logical operators for these boundary conditions is that the logical $\lx$ operator is a string of $X$ errors running from the top to the bottom boundary, and the logical $\lz$ operator is a string of $Z$ errors running from the left to the right boundary.
	\subsection{Noise models}
		While many types of noise are possible on quantum computers, here we focus on the restricted class of noise models called \textbf{independent non-identically distributed Pauli noise}, originally introduced in \cite{inidNoiseSurfaceCode}. In this class of noise models, every qubit $j$ has a Pauli $X$ error with probability $p_X^{(j)}$, a Pauli $Y$ error with probability $p_Y^{(j)}$, a Pauli $Z$ error with probability $p_Z^{(j)}$, and no error with probability $1-p_X^{(j)}-p_Y^{(j)}-p_Z^{(j)}$. Errors for different qubits are independent. The probabilities $p_X^{(j)}$, $p_Y^{(j)}$, $p_Z^{(j)}$ may be different for each qubit $j$. Equivalently, we may think of each qubit $j$ as undergoing the quantum channel,
		\begin{align*}
			\mathcal E^{(j)}(\rho)=(1-p_X^{(j)}-p_Y^{(j)}-p_Z^{(j)})\rho
			+p_X^{(j)} X\rho X + p_Y^{(j)} Y\rho Y + p_Z^{(j)} Z\rho Z.
		\end{align*}
	
		When we refer to a noise model, we mean a list of probabilities $p_X^{(j)}$, $p_Y^{(j)}$, $p_Z^{(j)}$ for each qubit $j$. Because we are restricting ourselves to this class of noise models, all errors we consider in this paper will be Pauli operators.
	\subsection{Non-degenerate versus degenerate quantum maximum likelihood decoding (QMLD versus DQMLD)}
		\label{QMLDvsDQMLD}
		In classical error correction, the decoding task is to find an error (ie, a set of bits that were flipped) that is consistent with the received corrupted codeword, then reverse that error by flipping those same bits. Additionally, we want to find such an error that has maximum probability of occurring, over all possible errors consistent with the received corrupted codeword. The analogous decoding strategy with quantum error correction is, given a set of syndromes and a noise model (\ie an algorithm to compute probabilities of errors), find the highest probability error $E$ consistent with those syndromes, then reverse that error by applying $E$ (since every Pauli is its own inverse). This decoding strategy is known as quantum maximum likelihood decoding (QMLD).
		
		\vspace{11pt}
		
		\defbox{
			\begin{restatable}{definition}{qmld} \textbf{Quantum maximum likelihood decoding (QMLD) for the surface code.}
				\label{qmld}
				
				The input to the computational problem is:
				\begin{itemize}
					\item A surface code instance, specified by the width $w$ and height $h$ of the grid of qubits over which the code is defined.
					\item Syndromes ($+1$ or $-1$) for each of the $wh-1$ stabilizer generators.
					\item A Pauli noise model, in the form of rational number Pauli error probabilities $p_X^{(j)}, p_Y^{(j)}, p_Z^{(j)}$ for each qubit $j$.
				\end{itemize}
				The output to the computational problem is a Pauli error $E$ over all of the $wh$ qubits such that:
				\begin{itemize}
					\item $E$ is consistent with the syndromes, meaning that $E$ commutes with all stabilizer generators that have a $+1$ measurement result and anticommutes with all stabilizer generators that have a $-1$ measurement result.
					\item $E$ has maximum probability over all such Pauli errors consistent with the syndromes. The probability of a Pauli error is calculated using the error probabilities $p_X^{(j)}, p_Y^{(j)}, p_Z^{(j)}$.
				\end{itemize}
			\end{restatable}
		}
	
		In general, there may be multiple correct answers for QMLD, because multiple different Pauli errors could have the same probability. The task of QMLD is just to output one of the correct answers.
		
		In the rest of this paper, when we say QMLD, we are specifically referring to the above problem, QMLD for the surface code.
	
		Quantum maximum likelihood decoding is also sometimes referred to as non-degenerate quantum maximum likelihood decoding, in contrast to degenerate quantum maximum likelihood decoding (defined below).
		
		Maximizing the probability of that single error $E$ ignores the fact that there are many errors other than $E$ that differ from $E$ only by a stabilizer, and thus are also corrected by $E$, a property of quantum codes called \emph{degeneracy}. Specifically, given a quantum code with stabilizer group $\mathcal S$, for any error $E$, the coset of the stabilizer group $E\mathcal{S}$ is the set of all errors that differ from $E$ only by a stabilizer and thus can be corrected by $E$:
		\begin{align*}
			E\mathcal S=\{ES \,\,\, | \,\,\, S\in\mathcal S\}.
		\end{align*}
		We say that all errors in a coset $E\mathcal S$ are \emph{logically equivalent}, because all such errors differ by only a stabilizer, \ie a logical identity operator. Instead of finding a single error $E$ with maximum probability, a better strategy is to find an error $E$ such that the coset of logically equivalent errors $E\mathcal{S}$ has maximum sum of probabilities. That is, we want to maximize
		\begin{align}
			\pr{E\mathcal S}=\sum_{E'\in E\mathcal S}\pr{E'}\label{cosetProbability}
		\end{align}
		where $\pr{E'}$ is the probability of the Pauli error $E'$ occurring. \eqref{cosetProbability} is called the \textbf{coset probability}. The task of degenerate quantum maximum likelihood decoding (DQMLD) is to find an error $E$ that maximizes that coset probability $\pr{E\mathcal S}$.
		
		\vspace{11pt}
		
		\defbox{
			\begin{restatable}{definition}{dqmld} \textbf{Degenerate quantum maximum likelihood decoding (DQMLD) for the surface code.}
				\label{dqmld}
				
				The input to the computational problem (identical to that of the non-degenerate problem) is:
				\begin{itemize}
					\item A surface code instance, specified by the width $w$ and height $h$ of the grid of qubits over which the code is defined.
					\item Syndromes ($+1$ or $-1$) for each of the $wh-1$ stabilizer generators.
					\item A Pauli noise model, in the form of rational number Pauli error probabilities $p_X^{(j)}, p_Y^{(j)}, p_Z^{(j)}$ for each qubit $j$.
				\end{itemize}
				The output to the computational problem is a Pauli error $E$ over all of the $wh$ qubits such that:
				\begin{itemize}
					\item $E$ is consistent with the syndromes, meaning that $E$ commutes with all stabilizer generators that have a $+1$ measurement result and anticommutes with all stabilizer generators that have a $-1$ measurement result.
					\item $E$ has maximum coset probability,
					\begin{align*}
						\sum_{E'\in E\mathcal S}\pr{E'},
					\end{align*}
					over all such Pauli errors consistent with the syndromes. Here $\mathcal S$ is the stabilizer group for the surface code on the $w\times h$ grid of qubits.
				\end{itemize}
			\end{restatable}
		}
	
		Note that there are always multiple correct answers for DQMLD, because for any Pauli error $E$, every other Pauli error in $E\mathcal S$ (the set of logically equivalent Paulis) has the same coset probability. The task of DQMLD is to just output one of the correct answers. Again, in the rest of this paper, when we say DQMLD, we are specifically referring to the above problem, DQMLD for the surface code. The difference between definitions \ref{qmld} and \ref{dqmld} is that \defref{qmld} maximizes the probability of the error found while \defref{dqmld} maximizes the coset probability of the error found. The terminology about these two decoding problems is somewhat inconsistent in the literature. DQMLD is sometimes referred to simply as maximum likelihood (ML) decoding (\eg in \cite{BSV14}). QMLD is sometimes referred to as maximum probability (MP) decoding (\eg in \cite{ChubbFlammiaStatMechSurfaceCode}). Here we use the terminology of \cite{surfaceCodeDecodersReview23}.
		
		To further understand the difference between QMLD and DQMLD, we can consider the $TLS$ decomposition of Paulis. Given a quantum code on $n$ qubits, one can associate to any possible set of syndromes $s$ an $n$-qubit Pauli operator $T(s)$ consistent with the syndromes $s$. This function $T$ partitions the $n$-qubit Paulis into sets of errors that all cause the same syndromes $s$, as we can write any $n$-qubit Pauli that causes the syndromes $s$ as $T(s)N$, where $N$ is a Pauli that commutes with all stabilizers. QMLD can be thought of as finding an $N$ that commutes with all stabilizers and maximizes the probability of the error $T(s)N$, given the syndromes $s$.
		
		However, we can decompose $N$ further. Any such $N$ that commutes with all stabilizers is a member of $\mathcal N(\mathcal S)$, the normalizer of the stabilizer group $\mathcal S$. Because $\mathcal N(\mathcal S)$ is generated by $\mathcal S$ and the logical operators of the code, we can write any such $N\in\mathcal N(\mathcal S)$ as $LS$, where $L$ is a logical operator and $S$ is a stabilizer. For the surface code, there are only 4 logical operators: $\li$, $\lx$, $\ly$, $\lz$. Thus any Pauli that causes syndromes $s$ can be written as $T(s)LS$, where $L\in\{\li, \lx, \ly, \lz\}$. Because changing $S$ to a different stabilizer does not affect the state, we don't care about recovering $S$, but rather we must pick $L$ correctly, in a way that maximizes the coset probability $\sum_{S\in\mathcal S}\pr{T(s)LS}$.
		
		To summarize, we can write any Pauli error $P$ that causes syndromes $s$ as
		\begin{align*}
			P=T(s)N=T(s)LS
		\end{align*}
		where $N=LS$ is a Pauli that commutes with all stabilizers, $L$ is one of the 4 logical operators for the surface code, and $S$ is a stabilizer. The syndrome measurements $s$ fix $T(s)$. QMLD seeks to find an $N$ that commutes with all stabilizers and maximizes the probability of the resulting Pauli error. DQMLD seeks to pick a logical operator $L\in\{\li, \lx, \ly, \lz\}$ that maximizes the sum of error probabilities for all $S\in\mathcal S$.
		
		DQMLD is the better decoding strategy because it maximizes the probability that no logical error occurs after we perform the correction, which is what we really care about. However, DQMLD can be less computationally feasible because exactly solving it can involve summing over exponentially many error probabilities (for exponentially many elements of $\mathcal S$).
		
		Both QMLD and DQMLD for the surface code are known to be exactly solvable in the case of independent $X$ and $Z$ noise. In the case of QMLD, the standard matching decoder\cite{topologicalQuantumMemory} finds the maximum probability error by finding separately the minimum weight matching for the $X$ syndromes and for the $Z$ syndromes and combining those matchings into one error. DQMLD can be reduced to matchgate quantum circuit simulation in the case of independent $X$ and $Z$ noise (\cite{BSV14} Section V).
		
		Our reduction outputs noise models with explicitly non-independent $X$ and $Z$ noise. This means that the matching decoder\cite{topologicalQuantumMemory} and the matchgate-circuit-simulation-based decoder\cite{BSV14} are not known to exactly solve the QMLD or DQMLD problems for the problem instances outputted by our reduction.

\pagebreak
\section{Reduction from SAT to quantum maximum likelihood decoding (QMLD)} \label{qmldHardSection}
	In this section we prove our first main theorem.
	\qmldHardTheorem
	
	\defbox{\qmld*}
	
	We reduce directly from SAT. Before we give the reduction from SAT to QMLD for the surface code, we review some facts about SAT and explain our graphical notation for writing down noise models.
	
	SAT (boolean formula satisfiability) is the problem of determining whether a boolean formula made of AND, OR, and NOT gates has an assignment of true/false values to the variables that makes the formula output true. For example, the boolean formula $(x_1\vee x_2)\wedge(\bar x_2\vee x_3)\wedge(\bar x_1\vee \bar x_3)$\footnote{Recall $\wedge=\andd$, $\vee=\orr$, $\bar x=\nott(x)$.} is satisfiable because one can assign the input variables the values $x_1=\true$, $x_2=\false$, $x_3=\false$. SAT is the canonical NP-complete problem\cite{cook71complexityTheoremProvingProcedures}. This means that all problems in the class NP can be efficiently transformed into a SAT problem, and thus an efficient algorithm for SAT implies efficient algorithms for every problem in NP.
	
\begin{figure*}
		\centering\begin{tikzpicture}[x=0.8cm,y=0.8cm]
			\node at (-3, 0) {input wires:};
			\node at (0, 0) {$x_1$};
			\node at (2, 0) {$x_2$};
			\node at (4, 0) {$\cdots$};
			\node at (6, 0) {$x_n$};
			
			\draw (0, 0.3) -- (0, 1);
			\node at (0, 1) [fill,circle,inner sep=2pt] {};
			\draw (0, 1) -- (-1,1) -- (-1, 4);
			\draw (0, 1) -- (3,1) -- (3, 4);
			
			\draw (2, 0.3) -- (2, 0.9);
			\draw (2, 1.1) -- (2, 2);
			\node at (2, 2) [fill,circle,inner sep=2pt] {};
			\draw (2, 2) -- (0,2) -- (0,4);
			\draw (1, 2) -- (1,4);
			\draw (2, 2) -- (2.9, 2);
			\draw (3.1, 2) -- (5, 2) -- (5, 4);
			
			\draw (6, 0.3) -- (6, 3);
			\node at (6, 3) [fill,circle,inner sep=2pt] {};
			\draw (6, 3) -- (5.1, 3);
			\draw (4.9, 3) -- (3.1, 3);
			\draw (2.9, 3) -- (2, 3);
			\draw (2, 3) -- (2, 4);
			\draw (4, 3) -- (4, 4);
			\draw (6, 3) -- (6, 4);
			
			\node[align=left] at (-3.5, 3) {FAN-OUT gates + \\ wire crossings:};
			
			\node [align=center] at (2.5, 5) {AND, OR, and NOT gates \\ (without wire crossings)};
			
			\draw[draw=black,line width=1pt] (-1.5,4) rectangle ++(8,2);
			
			\draw (2.5, 6) -- (2.5, 7.25);
			\node at (2.5, 7.5) {output wire};
			
			\draw (9, 1) -- (9, 2);
			\node at (9, 2) [fill,circle,inner sep=2pt] {};
			\draw (9, 2) -- (8, 2) -- (8, 3);
			\draw (9, 2) -- (9, 3);
			\draw (9, 2) -- (10, 2) -- (10, 3);
			\node at (11.5, 2) {= FAN-OUT};
		\end{tikzpicture}
		\caption{Any boolean formula can be viewed as a boolean circuit with a layer of FAN-OUT gates on the input wires, then some wire crossings, then AND, OR, and NOT gates without any wire crossings. This is because the part of the circuit graph above the FAN-OUT gates is a tree, and all trees are planar.}\label{booleanCircuit}
	\end{figure*}
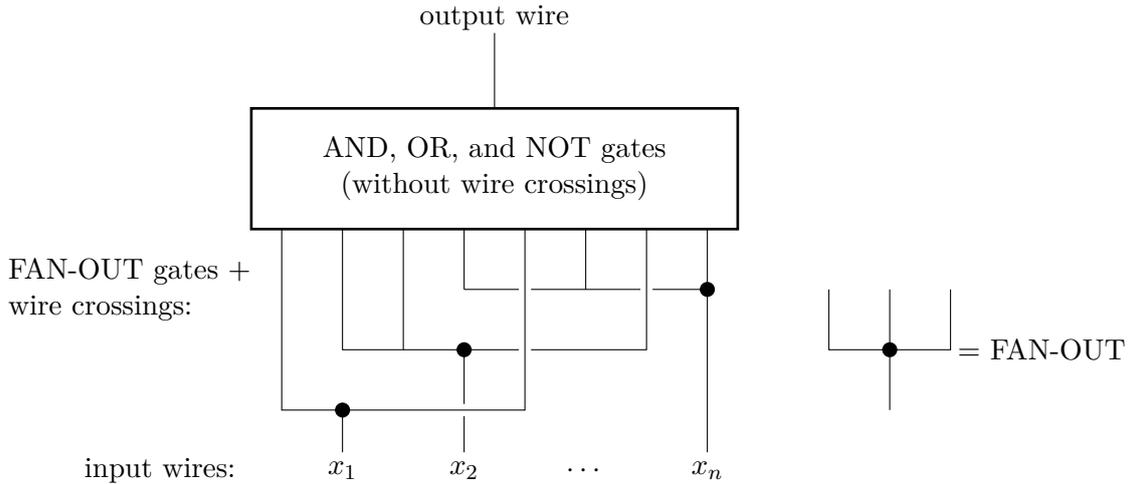

	We can think of any boolean formula as a boolean circuit consisting of a layer of FAN-OUT gates (gates that make copies of their inputs) after the input wires, then the relevant AND, OR, and NOT gates after that layer. Without loss of generality, we can assume that no wires cross each other after the layer of FAN-OUT gates; this is because the circuit graph of the part of the circuit after the FAN-OUT gates is a tree (it is the parse tree of the boolean formula), and all trees are planar. See \figref{booleanCircuit} for an example. The planarity of this part of the circuit is important because we will have to carefully argue how we can handle the wire crossings after the FAN-OUT gates.
	
	Now we describe our notation for writing down noise models and errors. We consider Pauli noise models where every qubit $j$ has specific probabilities $p^{(j)}_X$, $p^{(j)}_Y$, $p^{(j)}_Z$ of $X$, $Y$, and $Z$ errors respectively. In all the figures in this paper, drawing red (bold-faced) operators on a qubit means we are writing down a specific Pauli error. Drawing purple (non-bold) operators on a qubit means we are writing down a noise model---that is, the list of probabilities $p^{(j)}_X$, $p^{(j)}_Y$, $p^{(j)}_Z$ for each qubit. If we draw a single operator (\eg $X$) on a qubit when describing a noise model, that means that that error occurs on that qubit with the fixed probability $p$, and no error occurs with probability $1-p$. Our results hold if $p$ is any fixed constant in $(0, 0.25]$. If we draw multiple operators on a qubit (\eg if we draw $X,Z$, or if we draw $X,Y,Z$), that means that each of those operators can occur as an error, each with probability $p$, and no error occurs with probability $1-2p$ or $1-3p$. If we draw a noise model on a surface code without a boundary, it is to be understood that it is only a subset of the whole surface code patch and will be stitched together with the other gadgets.
	
	\begin{figure}
		\begin{subfigure}{0.5\textwidth}
			\centering\begin{tikzpicture}[x=0.5cm,y=0.5cm]
				\surfaceCodeCheckerboard{5}{5}{false}{false}{true}
				\node at (0, 1) [noiseStyle] {X};
				\node at (1, 1) [noiseStyle] {Y,Z};
				\node at (3, 1) [noiseStyle] {X,Y,Z};
				
				\node at (0, 3) [noiseStyle] {Z};
				\node at (1, 3) [noiseStyle] {Y,X};
				\node at (3, 3) [noiseStyle] {X,Y,Z};
			\end{tikzpicture}
			\caption{}\label{exampleNoiseModel}
		\end{subfigure}
		\begin{subfigure}{0.5\textwidth}
			\centering\begin{tikzpicture}[x=0.5cm,y=0.5cm]
				\surfaceCodeCheckerboard{5}{5}{false}{false}{true}
				\node at (0, 1) [noiseStyle] {X};
				\node at (1, 1) [noiseStyle] {Y,Z};
				\node at (3, 1) [noiseStyle] {X,Y,Z};
				
				\node at (0, 3) [noiseStyle] {Z};
				\node at (1, 3) [noiseStyle] {Y,X};
				\node at (3, 3) [noiseStyle] {X,Y,Z};
				
				\node at (0, 1) [errorStyle] {X};
				
				\node at (1, 1) [errorStyle] {Y};
				
				\node at (3, 1) [errorStyle] {X};
			\end{tikzpicture}
			\caption{}\label{exampleErrorNoiseModel}
		\end{subfigure}
		\caption{\subref{exampleNoiseModel} An example of our graphical notation for writing down noise models, which in this paper are just probability distributions of Pauli errors. For every operator drawn on a qubit, that error occurs on that qubit with the probability $p$, which we can choose to be any fixed constant in $(0,0.25]$. If no operators are drawn on a qubit, then errors have 0 probability for that qubit. The errors for different qubits are independent. \subref{exampleErrorNoiseModel} An example of an error that could occur with this noise model. We denote errors with bold red letters, and often draw them on top of the noise model, as done here. This error occurs with probability $p^3(1-p)(1-2p)(1-3p)$. Here we do not draw the syndromes that result from this error.}
	\end{figure}
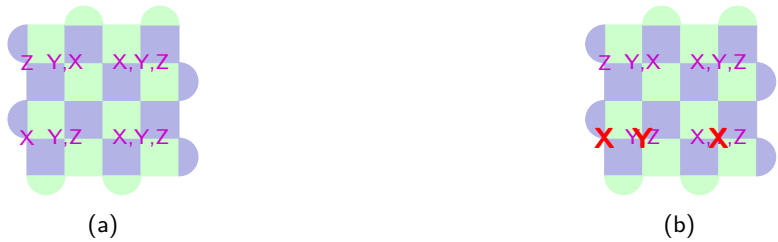

	We give an example of a noise model using our graphical notation in \figref{exampleNoiseModel}. There are 6 qubits with 1, 2, or 3 Pauli errors possible, each with probability $p$, and all other qubits have 0 probability of error. \figref{exampleErrorNoiseModel} shows one possible error that could result from this noise model (we do not draw the syndromes caused by this error). There we draw the error with red, bold-faced letters, on top of the noise model, which we often do to make it easier to visualize what possible errors could arise from a noise model. That error occurs with probability $p^3(1-p)(1-2p)(1-3p)$. This is because there are 3 qubits with an error (each of which happens with probability $p$), one qubit with one possible error that does not have an error occur (this occurs with probability $1-p$), one qubit with two possible errors that does not have an error occur (this occurs with probability $1-2p$), and one qubit with three possible errors that does not have an error occur (this occurs with probability $1-3p$).
	
	\subsection{Overview of the reduction}
		\label{overviewReductionSection}
		
		\begin{figure}
			\centering\begin{tikzpicture}
				\node[align=left,font=\small] at (0.2, 0) {input: \\ boolean \\ formula};
				
				\draw[->] (1.1, 0) -- (1.9, 0);
				
				\draw[draw=black,line width=1pt,font=\small] (2.1, -0.8) rectangle ++(2.6,1.6);
				\node[font=\small,align=left] at (3.4, 0) {SAT to QMLD \\ conversion};
				\node[font=\small,align=left] at (3.4, 1.1) {(a)};
				
				\draw[->] (4.9, 0) -- (5.7, 0);
				
				\node[align=left,font=\small] at (7.3, 0) {QMLD problem \\ instance: \\ $\cdot$ surface code size\\$\cdot\,\,p_X^{(j)},p_Y^{(j)},p_Z^{(j)}$\\ $\cdot$ syndromes};

				\draw [->] (8.8, 0) -- (9.6, 0);
				
				\draw[draw=black,line width=1pt,font=\small] (9.8, -0.8) rectangle ++(2.6,1.6);
				\node[font=\small,align=left] at (11.1, 0) {QMLD \\ oracle};
				\node[font=\small,align=left] at (11.1, 1.1) {(b)};
				
				\draw (11.1, -1.0) -- (11.1, -1.5) -- (2.1, -1.5) -- (2.1, -3.0);
				\draw [->] (2.1, -3.0) -- (3.1, -3.0);
				
				\node[align=left,font=\small] at (4.4, -3.0) {Pauli error $E$ \\ (requirements \\ in Def.~\ref{qmld})};
				
				\draw [->] (5.8, -3.0) -- (6.7, -3.0);
				
				\draw[draw=black,line width=1pt,font=\small] (6.9, -3.8) rectangle ++(2.6,1.6);
				\node[font=\small,align=left] at (8.2, -3.0) {algorithm that \\ examines $E$};
				\node[font=\small,align=left] at (9.8, -3.0) {(c)};
				
				\draw [->] (8.2, -4.0) -- (8.2, -5.0);
				
				\node[font=\small,align=center] at (8.2, -5.5) {output: \\ satisfiable/unsatisfiable};
				
				\draw (8.3, -4.5) -- (13.0, -4.5) -- (13.0, 1.9) -- (1.5, 1.9) -- (1.5, 0.1);
				\draw (1.5, -0.1) -- (1.5, -4.5) -- (8.1, -4.5);
				\node[font=\small,align=left] at (11.1, -4.8) {(d)};
				
			\end{tikzpicture}
			\caption{Diagram of our Turing reduction from SAT to QMLD. In Section \ref{qmldHardSection} we give a polynomial-time algorithm that converts a boolean formula to a QMLD problem instance (surface code size, list of error probabilities, and syndromes); this algorithm is (a). A hypothetical algorithm that exactly solves QMLD would produce a Pauli error $E$ when given that QMLD problem as its input; this hypothetical algorithm is (b), the QMLD oracle. Given that error $E$, one can determine in polynomial time whether the original boolean formula is satisfiable; the algorithm that does this is (c) and is also described in Section 3. This means that the existence of an algorithm that exactly solves QMLD in polynomial time implies the existence of an algorithm that solves SAT in polynomial time; this algorithm is (d). This establishes QMLD as NP-hard, via a Turing reduction.}
			\label{reductionSequence}
		\end{figure}
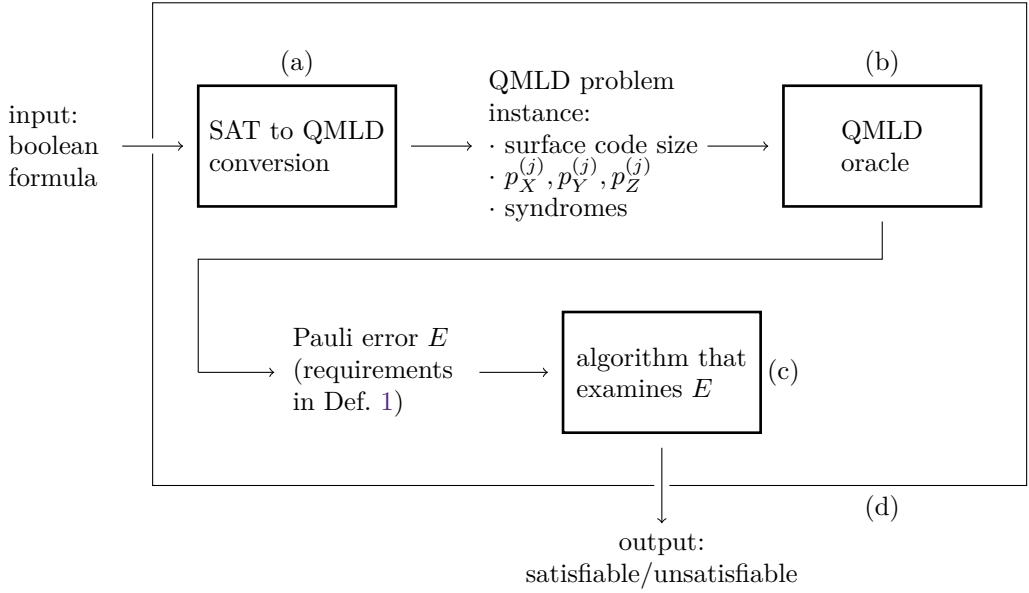
		
		Our reduction is a Turing reduction that uses a QMLD oracle to solve SAT in polynomial time. There are two parts to our Turing reduction. The first part is a polynomial-time algorithm that takes as input a SAT instance with $n$ variables, and outputs a QMLD problem instance (surface code size, noise model, and set of syndromes) with the following properties:
		\begin{enumerate}
			\item There are exactly $2^n$ Pauli errors with nonzero probability consistent with the given syndromes: one for each setting of the $n$ variables of the boolean formula.\label{propertyErrorsAssignments}
			\item For any Pauli error with nonzero probability consistent with the given syndromes, it is easy (\ie can be done in polynomial time) to tell whether that error corresponds to a satisfying or unsatisfying assignment of the variables of the boolean formula.\label{propertyTellSatisfying}
			\item The noise model is such that the maximum probability error consistent with the given syndromes is always an error corresponding to a satisfying assignment of the variables of the boolean formula, if such an assignment exists.\label{propertyRigged}
		\end{enumerate}
	
		The second part of our Turing reduction is the algorithm referenced in property \ref{propertyTellSatisfying} above: it is the polynomial-time algorithm that takes as input the Pauli error $E$ that is the answer to the QMLD problem, and outputs satisfiable or unsatisfiable. \figref{reductionSequence} gives a diagrammatic overview of the steps in our Turing reduction; there box (a) is the first part of our reduction (conversion from boolean formula to QMLD problem instance) and box (c) is the second part of our reduction (algorithm that takes Pauli error $E$ as input and outputs satisfiable/unsatisfiable).
	
		Composing the first part of our reduction (conversion from boolean formula to QMLD problem instance, box (a) in \figref{reductionSequence}) with an oracle for QMLD (box (b) in \figref{reductionSequence}) then with the algorithm that examines the Pauli error $E$ (box (c) in \figref{reductionSequence}) gives a polynomial-time algorithm that solves SAT; this composition is our Turing reduction from SAT to QMLD. \figref{reductionSequence} shows box (d) as the composition between boxes (a), (b), and (c), that together form the polynomial-time algorithm that solves SAT. The second part of our reduction is quite simple, and the first part is the complicated part that most all of Section \ref{qmldHardSection} is devoted to describing.
		
		
		Here we only concern ourselves with Turing reductions (\ie using an oracle for QMLD for the surface code to solve SAT in polynomial time), rather than many-one/Karp reductions (\ie transforming a SAT instance into a decoding problem instance with the same yes/no answer). This is because QMLD is most naturally thought of as a function problem (outputting a Pauli error), rather than a yes/no decision problem. In appendix \ref{completenessAppendix}, we show that QMLD becomes NP-complete if we reformulate it as a decision problem, but in the main text we choose to only think about the more natural function problem version of QMLD, for clarity.
	
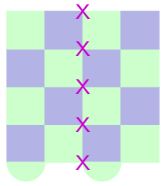
\begin{figure*}
			\centering\begin{tikzpicture}[x=0.5cm,y=0.5cm]
				\surfaceCodeCheckerboard{5}{5}{false}{false}{false}
				
				\foreach \x in {0,2} {
					\filldraw[draw=xStabColor,fill=xStabColor]
					(\x, 0) arc (-180:0:0.5) -- cycle;
				}
				
				\foreach \y in {0,...,4} \node at (2, \y) [noiseStyle] {X};
				
			\end{tikzpicture}
			\caption{The variable gadget. The noise model and lack of $-1$ syndromes given means that either there is one string of $X$ errors starting at the bottom boundary and continuing up (which corresponds to the variable being true), or there are no errors (which corresponds to the variable being false).}\label{variableGadget}
		\end{figure*}

		With this high-level overview of the reduction in mind, we can now define the first gadget in our conversion from a boolean formula to a QMLD problem instance, the variable gadget, which is in \figref{variableGadget}. In this gadget the noise model and lack of $-1$ syndromes mean there will either be one string of $X$ errors starting at the bottom boundary and continuing up to the rest of the construction, or no errors at all in this gadget. If the string of $X$ errors goes up to the rest of the circuit, then that variable is true, and if there is no such string of $X$ errors, then that variable is false. In general a ``wire'' in this reduction will be a string of possible $X$ errors---if there actually are $X$ errors present in that string in the error, then that wire is true, otherwise that wire is false.
		
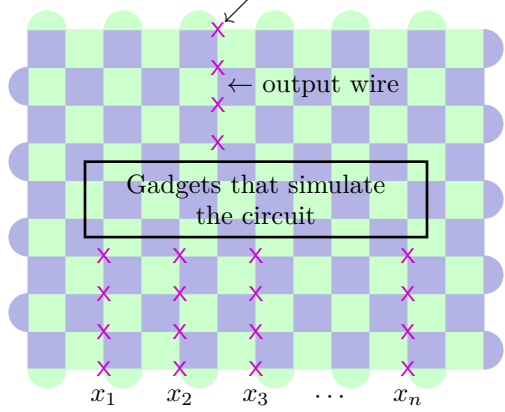
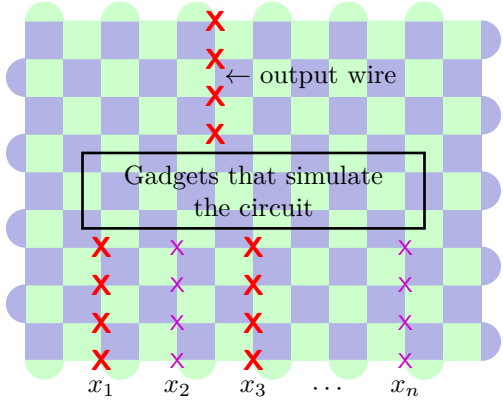
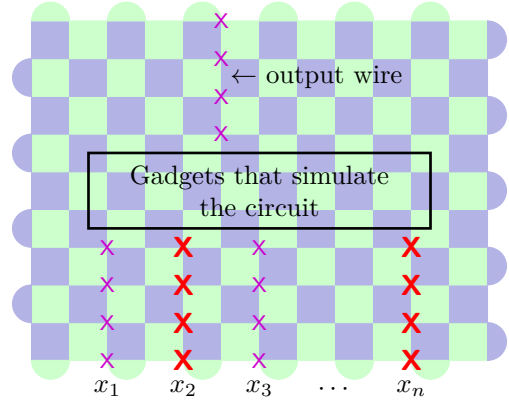
\begin{figure*}
			\begin{subfigure}{\textwidth}
				\centering\begin{tikzpicture}[x=0.5cm,y=0.5cm]
					\surfaceCodeCheckerboard{13}{10}{false}{false}{true}
					
					\foreach \x in {2, 4, 6, 10} {
						\foreach \y in {0,...,3} {
							\node at (\x, \y) [noiseStyle] {X};
						}
					}
					\foreach \x in {1,2,3} {
						\node at (2*\x, -0.75) [labelStyle] {$x_\x$};
					}
					
					\node at (8, -0.75) [labelStyle] {$\cdots$};
					\node at (10, -0.75) [labelStyle] {$x_n$};
					
					\draw[draw=black,line width=1pt] (1.5,3.5) rectangle ++(9,2);
					\node at (6, 4.5) [labelStyle] {Gadgets that simulate \\ the circuit};
					
					\foreach \y in {6,...,9} \node at (5, \y) [noiseStyle] {X};
					\node at (7.5, 7.5) [labelStyle] {$\leftarrow$ output wire};
					\node at (5.5, 10) [labelStyle] {this $X$ error occurs with probability $1-p^{\ell}$, where $\ell=wh=$ \# of qubits in the code \\ $\swarrow$};
				\end{tikzpicture}
				\caption{Overview of the noise model and syndromes outputted by the conversion from a boolean formula to a QMLD problem instance. In general there will be larger spacing between the variable gadgets than the finite spacing shown here. The ``gadgets that simulate the circuit'' are AND, OR, NOT, and FAN-OUT gadgets that express any boolean formula as in \figref{booleanCircuit}.}\label{reductionOverview}
				\vspace{0.25cm}
			\end{subfigure}
		
			\begin{subfigure}{0.45\textwidth}
				\centering\begin{tikzpicture}[x=0.5cm,y=0.5cm]
					\surfaceCodeCheckerboard{13}{10}{false}{false}{true}
					
					\foreach \x in {2, 4, 6, 10} {
						\foreach \y in {0,...,3} {
							\node at (\x, \y) [noiseStyle] {X};
						}
					}
					\foreach \x in {1,2,3} {
						\node at (2*\x, -0.75) [labelStyle] {$x_\x$};
					}
					
					\node at (8, -0.75) [labelStyle] {$\cdots$};
					\node at (10, -0.75) [labelStyle] {$x_n$};
					
					\draw[draw=black,line width=1pt] (1.5,3.5) rectangle ++(9,2);
					\node at (6, 4.5) [labelStyle] {Gadgets that simulate \\ the circuit};
					
					\foreach \y in {6,...,9} \node at (5, \y) [noiseStyle] {X};
					\node at (7.5, 7.5) [labelStyle] {$\leftarrow$ output wire};
					
					\foreach \x in {2,6}
						\foreach \y in {0,...,3} \node at (\x, \y) [errorStyle] {X};
					
					\foreach \y in {6,...,9} \node at (5, \y) [errorStyle] {X};
					
				\end{tikzpicture}
				\caption{A hypothetical error corresponding to a satisfying assignment of the formula. If the presence of $X$ errors in the variable gadgets is chosen in a way that corresponds to choosing values of the variables so they satisfy the formula, then the output wire will have $X$ errors present.}\label{reductionOverviewSatisfyingAssignment}
			\end{subfigure}
			\hspace{0.1\textwidth}
			\begin{subfigure}{0.45\textwidth}
				\centering\begin{tikzpicture}[x=0.5cm,y=0.5cm]
					\surfaceCodeCheckerboard{13}{10}{false}{false}{true}
					
					\foreach \x in {2, 4, 6, 10} {
						\foreach \y in {0,...,3} {
							\node at (\x, \y) [noiseStyle] {X};
						}
					}
					\foreach \x in {1,2,3} {
						\node at (2*\x, -0.75) [labelStyle] {$x_\x$};
					}
					
					\node at (8, -0.75) [labelStyle] {$\cdots$};
					\node at (10, -0.75) [labelStyle] {$x_n$};
					
					\draw[draw=black,line width=1pt] (1.5,3.5) rectangle ++(9,2);
					\node at (6, 4.5) [labelStyle] {Gadgets that simulate \\ the circuit};
					
					\foreach \y in {6,...,9} \node at (5, \y) [noiseStyle] {X};
					\node at (7.5, 7.5) [labelStyle] {$\leftarrow$ output wire};
					
					\foreach \x in {4,10}
					\foreach \y in {0,...,3} \node at (\x, \y) [errorStyle] {X};
				\end{tikzpicture}
				\caption{A hypothetical error corresponding to an unsatisfying assignment of the formula. If the presence of $X$ errors in the variable gadgets is chosen in a way that corresponds to choosing values of the variables so they don't satisfy the formula, then the output wire will not have $X$ errors present.}\label{reductionOverviewUnsatisfyingAssignment}
			\end{subfigure}
			\caption{Overview of the conversion from a boolean formula to a QMLD problem instance, with examples of errors corresponding to a satisfying assignment and an unsatisfying assignment. This conversion is box (a) in \figref{reductionSequence}. One can tell whether a Pauli error $E$ corresponds to a satisfying or unsatisfying assignment of the formula by looking for the presence of $X$ errors in $E$ in the output wire; this procedure is box (c) of \figref{reductionSequence}.}
		\end{figure*}

		With this initial gadget defined, we can now give a more detailed overview of the conversion from a boolean formula to a QMLD problem instance before diving in to the rest of the gadget constructions. See \figref{reductionOverview}---in that figure, the ``gadgets that simulate the circuit'' are AND, OR, NOT, and FAN-OUT gadgets that express the boolean formula as a circuit, as in \figref{booleanCircuit}. These gadgets ``simulate the circuit'' in the following sense:
		\begin{itemize}
			\item If the presence of $X$ errors in the variable gadgets is set corresponding to a satisfying assignment of the formula, then the only possible error consistent with the syndromes has $X$ errors in the output wire. \figref{reductionOverviewSatisfyingAssignment} is a hypothetical example of such an error.
			\item If the presence of $X$ errors in the variable gadgets is set corresponding to an unsatisfying assignment of the formula, then the only possible error consistent with the syndromes has no $X$ errors present in the output wire. \figref{reductionOverviewUnsatisfyingAssignment} is a hypothetical example of such an error.
		\end{itemize}
		We give a special probability to the topmost possible $X$ error in the output wire of the circuit. We set this probability such that that error appearing is so likely (and it not appearing is so unlikely) that it forces the maximum probability error to correspond to a satisfying assignment of the circuit, if such a satisfying assignment exists.
		
		To make this notion precise, let $\ell=wh$ be the number of qubits in the surface code instance outputted by the reduction, and note that the probability of any error corresponding to a satisfying assignment of the circuit is lower bounded by
		\begin{align}
			\begin{split}
				& p^{\text{\# of qubits with an error, except topmost qubit of output wire}} \\
				\times & (1-p)^{\text{\# of qubits without an error, and 1 error was possible, except topmost qubit of output wire}} \\
				\times & (1-2p)^{\text{\# of qubits without an error, and 2 errors were possible}} \\
				\times & (1-3p)^{\text{\# of qubits without an error, and 3 errors were possible}} \\
				\times & \left(1-p^{\ell}\right) \\
				\geq & \left(1-p^{\ell}\right) p^{\ell-1} \\
				> & p^{\ell}.\label{riggedProbabilities}
			\end{split}
		\end{align}
		And the probability of any error corresponding to an unsatisfying assignment of the circuit is at most $p^{\ell}$, since the lack of presence of an $X$ error at the top of the output wire causes a $p^{\ell}$ term to appear in the probability of such an error. Therefore, any satisfying assignment of the circuit will correspond to an error with higher probability than any error corresponding to an unsatisfying assignment of the circuit. Therefore a QMLD oracle, when run on this QMLD problem instance, will always find an error corresponding to a satisfying assignment of the circuit, if one exists.
		
		It is easy to determine whether an error corresponds to a satisfying or unsatisfying assignment of the circuit by looking at the presence of $X$ errors in the output wire. This simple procedure gives us box (c) in \figref{reductionSequence} and satisfies property \ref{propertyTellSatisfying} listed above. Therefore, an oracle for QMLD for the surface code can be used to solve SAT in polynomial time via the procedure outlined in \figref{reductionSequence}, which establishes QMLD for the surface code as NP-hard.
		
		This establishes what ingredients we need to construct to complete the reduction. We need gadgets for AND, OR, NOT, and FAN-OUT gates, and for the wire crossings above the FAN-OUT gates. Recall that OR gates can be constructed from AND gates and NOT gates, because $\orr(x,y)=\nott(\andd(\nott(x),\nott(y)))$, so really all we need is AND, NOT, and FAN-OUT gates, and wire crossings. The remainder of this section is constructing these gadgets and arguing how they can be stitched together, in order to complete the conversion between a boolean formula and a QMLD problem instance, and thus complete the proof that QMLD for the surface code is NP-hard.
		
		Before we construct those gadgets, we construct 2 gadgets used as ingredients in the other gadgets. These are the ``convert to $Z$ ($X$) string'' and ``$X/Z$ wire crossing'' gadgets.
	\subsection{Convert between $X/Z$ string gadgets}
		\begin{figure*}
			\begin{subfigure}{0.5\textwidth}
				\centering\begin{tikzpicture}[x=0.5cm,y=0.5cm]
					\surfaceCodeCheckerboard{7}{9}{false}{false}{false}
					
					\foreach \y in {0,...,3} \node at (5, \y) [noiseStyle] {X};
					
					\node at (5, 4) [noiseStyle] {Y};
					
					\foreach \y in {5,...,8} \node at (5, \y) [noiseStyle] {Z};
					
					\node at (4, 4) [noiseStyle] {X};
					\node at (4, 3) [noiseStyle] {Z};
					\node at (3, 3) [noiseStyle] {Y};
					
					\foreach \y in {1,2}  \node at (3, \y) [noiseStyle] {X};
					
					\foreach \x in {1,2}  \node at (\x, 3) [noiseStyle] {Z};
					
					\node at (5, -1) [labelStyle] {$\uparrow$ \\ input $X$ string};
					\node at (5, 9) [labelStyle] {output $Z$ string \\ $\downarrow$};
					\node at (-0.3, 0) [labelStyle] {this string goes to $\nearrow$ \\ bottom $X$ boundary};
					\node at (-2.4, 3.3) [labelStyle] {this string goes to \\ left $Z$ boundary $\longrightarrow$};
				\end{tikzpicture}
				\caption{}\label{gadgetConvertX2Z}
			\end{subfigure}
			\begin{subfigure}{0.5\textwidth}
				\centering\begin{tikzpicture}[x=0.5cm,y=0.5cm]
					\surfaceCodeCheckerboard{8}{9}{false}{false}{false}
					
					\foreach \y in {0,...,3} \node at (3, \y) [noiseStyle] {Z};
					
					\node at (3, 4) [noiseStyle] {Y};
					\node at (4, 5) [noiseStyle] {Z};
					\node at (4, 6) [noiseStyle] {Z};
					\node at (3, 6) [noiseStyle] {Y};
					
					\foreach \y in {5,7,8} \node at (3, \y) [noiseStyle] {X};
					
					\node at (4, 4) [noiseStyle] {X};
					\node at (5, 4) [noiseStyle] {X};
					
					\foreach \y in {3,2}  \node at (6, \y) [noiseStyle] {X};
					
					\foreach \x in {2,1}  \node at (\x, 6) [noiseStyle] {Z};
					
					\node at (3, -1) [labelStyle] {$\uparrow$ \\ input $Z$ string};
					\node at (3, 9) [labelStyle] {output $X$ string \\ $\downarrow$};
					\node at (6, 0.5) [labelStyle] {$\downarrow$ \\ to bottom \\ boundary};
					\node at (-2.4, 6) [labelStyle] {to left boundary $\longleftarrow$};
				\end{tikzpicture}
				\caption{}\label{gadgetConvertZ2X}
			\end{subfigure}
			\caption{Two gadgets to convert between $X$ and $Z$ strings. \subref{gadgetConvertX2Z} converts its input $X$ string to a $Z$ string, and \subref{gadgetConvertZ2X} converts $Z$ to $X$. The only 2 Pauli errors with nonzero probability consistent with the syndromes are: either all the depicted errors are present, or none of the errors are present. These gadgets result in 2 other strings that will have to be routed to the appropriate boundary when stitching all the gadgets together.}\label{gadgetsConvertXZ}
		\end{figure*}
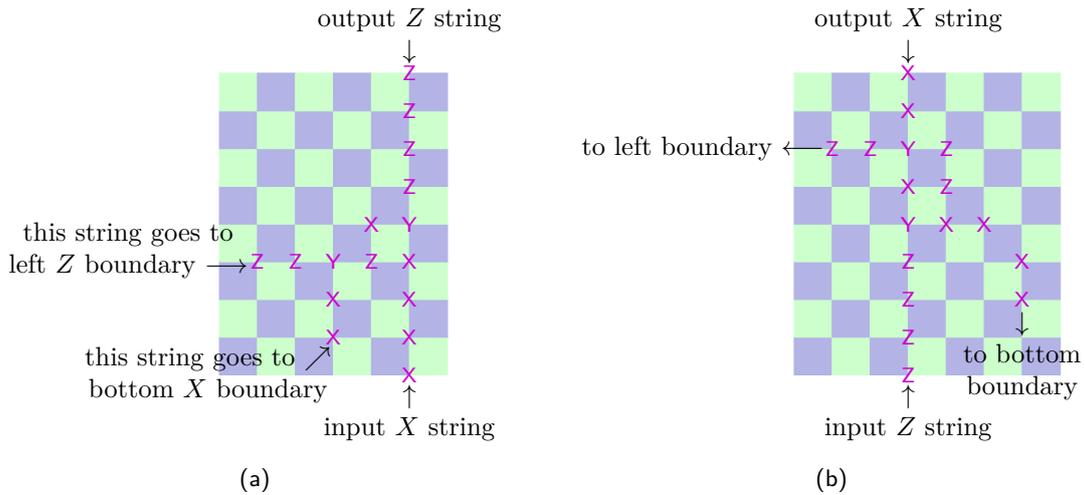
	
		Strings of $Z$ errors that act as wires in the circuit, instead of strings of $X$ errors, will be useful. So we have 2 gadgets to convert between them in \figref{gadgetsConvertXZ}. In these gadgets, the only 2 Pauli errors with nonzero probability consistent with the syndromes (all syndromes are $+1$) are: either all the depicted errors are present, or none of the errors are present. These gadgets result in an extra $X$ string and $Z$ string that have to be routed to the appropriate boundaries.
		
		Note that these gadgets contain \emph{non-independent $X$ and $Z$ error probabilities} for some qubits: specifically, the qubits where a $Y$ error is possible but not an $X$ nor $Z$ error. This non-independence of $X$ and $Z$ error probabilities means that the known decoders are not guaranteed to find the most likely error.
	\subsection{$X/Z$ wire crossing gadget}
		\begin{figure*}
			\centering\begin{tikzpicture}[x=0.5cm,y=0.5cm]
				\surfaceCodeCheckerboard{7}{7}{false}{false}{false}
				\foreach \y in {0,1,2,4,5,6}  \node at (3, \y) [noiseStyle] {X};
				\foreach \x in {0,1,2,4,5,6}  \node at (\x, 3) [noiseStyle] {Z};
				\node at (3,3) [noiseStyle] {X,Y,Z};
			\end{tikzpicture}
			\caption{Gadget that lets $X$ strings cross $Z$ strings.}\label{xzCrossGadget}
		\end{figure*}
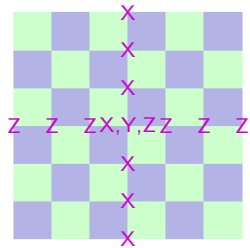
	
		\figref{xzCrossGadget} has a gadget that lets $X$ strings cross $Z$ strings. The 4 possible errors here are no error, just the $X$ string, just the $Z$ string, or both strings (with a $Y$ error at their intersection).
	
	\subsection{FAN-OUT gadget}
		\begin{figure*}
			\centering\begin{tikzpicture}[x=0.5cm,y=0.5cm]
				\surfaceCodeCheckerboard{19}{9}{false}{false}{false}
				
				\foreach \y in {0,1,2,4,6,7} \node at (7, \y) [noiseStyle] {X};
				\foreach \x in {8,9,10} \node at (\x, 7) [noiseStyle] {X};
				\foreach \y in {6,4,2} \node at (11, \y) [noiseStyle] {X};
				
				\node at (3,3) [noiseStyle] {Y};
				\node at (7,3) [noiseStyle] {Y};
				\node at (11,3) [noiseStyle] {Y};
				\node at (15,3) [noiseStyle] {Y};
				\node at (3,5) [noiseStyle] {Y};
				\node at (7,5) [noiseStyle] {Y};
				\node at (11,5) [noiseStyle] {Y};
				\node at (15,5) [noiseStyle] {Y};
				
				\foreach \y in {1,2,4,6,7,8} \node at (3, \y) [noiseStyle] {X};
				
				\foreach \y in {1,2,4,6,7,8} \node at (15, \y) [noiseStyle] {X};
				
				\foreach \x in {1,2,4,5,6,8,9,10,12,13,14,16} \node at (\x, 3) [noiseStyle] {Z};
				\node at (17,4) [noiseStyle] {Z};
				\foreach \x in {17,16,14,13,12,10,9,8,6,5,4,2,1} \node at (\x, 5) [noiseStyle] {Z};
				
				
				\node at (-2.2, 3) [labelStyle] {to left boundary $\longleftarrow$};
				\node at (-2.2, 5) [labelStyle] {to left boundary $\longleftarrow$};
				\node at (3, 0) [labelStyle] {$\downarrow$ \\ to bottom boundary};
				\node at (11, 1) [labelStyle] {$\downarrow$ \\ to bottom boundary};
				\node at (15, 0) [labelStyle] {$\downarrow$ \\ to bottom boundary};
				\node at (7,-1) [labelStyle] {$\uparrow$ \\ input wire};
				\node at (3,8.5) [labelStyle] {output 1};
				\node at (15,8.5) [labelStyle] {output 2};
			\end{tikzpicture}
			\caption{The FAN-OUT gadget. This gadget can be made arbitrarily wide by moving the input wire and output 1 to the left.}\label{fanoutGadget}
		\end{figure*}
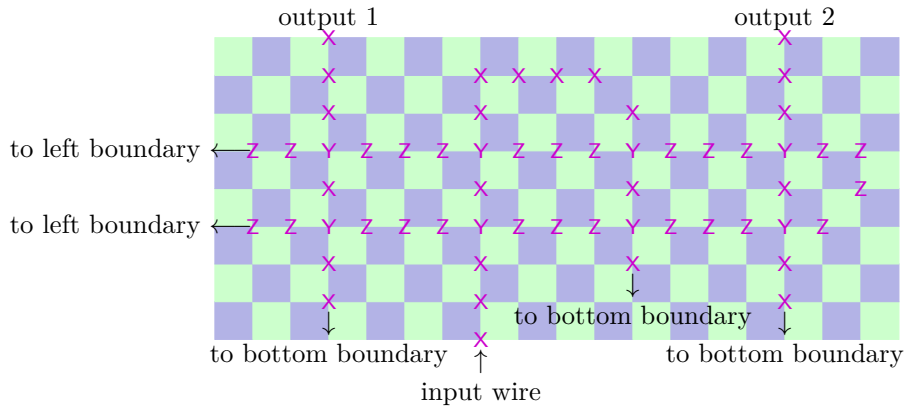
		
		The FAN-OUT gadget is in \figref{fanoutGadget}. This gadget can be made arbitrarily wide, instead of having the finite width as shown. This is necessary because the 2 output wires may have to be separated by a large horizontal distance, as in \figref{booleanCircuit}. In order to achieve the wire crossings above the FAN-OUT gates in \figref{booleanCircuit}, we only need to cross vertical wires (\ie $X$ wires) and horizontal wires (\ie $Z$ wires), which we can achieve with the $X/Z$ wire crossing gadget (\figref{xzCrossGadget}).
		
		This property of the FAN-OUT gadget, that $Z$ error strings are the only possible error strings that propagate horizontally across the whole width of the circuit and $X$ error strings are the only possible error strings that propagate vertically across the whole height of the circuit, will be shared across all the gadgets we construct. We already see this property in the ``convert to $Z$ string'' gadget (\figref{gadgetsConvertXZ}). That gadget has a possible string of $Z$ errors that goes horizontally all the way to the left boundary, and a possible string of $X$ errors that goes vertically all the way to the bottom boundary. When stitching all these gadgets together, these error strings can cross each other via the $X/Z$ wire crossing gadget (\figref{xzCrossGadget}).
	\subsection{NOT gadget}
		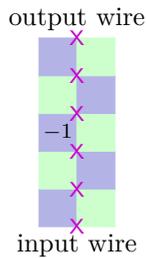
\begin{figure*}
			\centering\begin{tikzpicture}[x=0.5cm,y=0.5cm]
				\surfaceCodeCheckerboard{3}{6}{false}{false}{false}
				\foreach \y in {0,...,5}  \node at (1, \y) [noiseStyle] {X};
				\node at (0.5, 2.5) [syndromeStyle] {$-1$};
				\node at (1, -0.5) [labelStyle] {input wire};
				\node at (1, 5.5) [labelStyle] {output wire};
			\end{tikzpicture}
			\caption{The NOT gadget. If there is a string of $X$ errors coming up the input wire from below, that string will end at the $-1$ syndrome. If there is no string of $X$ errors coming up from below, then a string of $X$ errors will propagate upwards from the $-1$ syndrome, up the output wire.}\label{notGadget}
		\end{figure*}
	
		The NOT gadget is simple, just a $-1$ syndrome on a $Z$ stabilizer adjacent to the wire---see \figref{notGadget}.
		
		Clearly, if there is a string of $X$ errors going up from the bottom of the gadget, then there will be no string of $X$ errors going up above the $-1$ syndrome. And if there is no string of $X$ errors going up from the bottom of the gadget, then there will be a string of $X$ errors going up above the $-1$ syndrome.
	\subsection{AND gadget}
		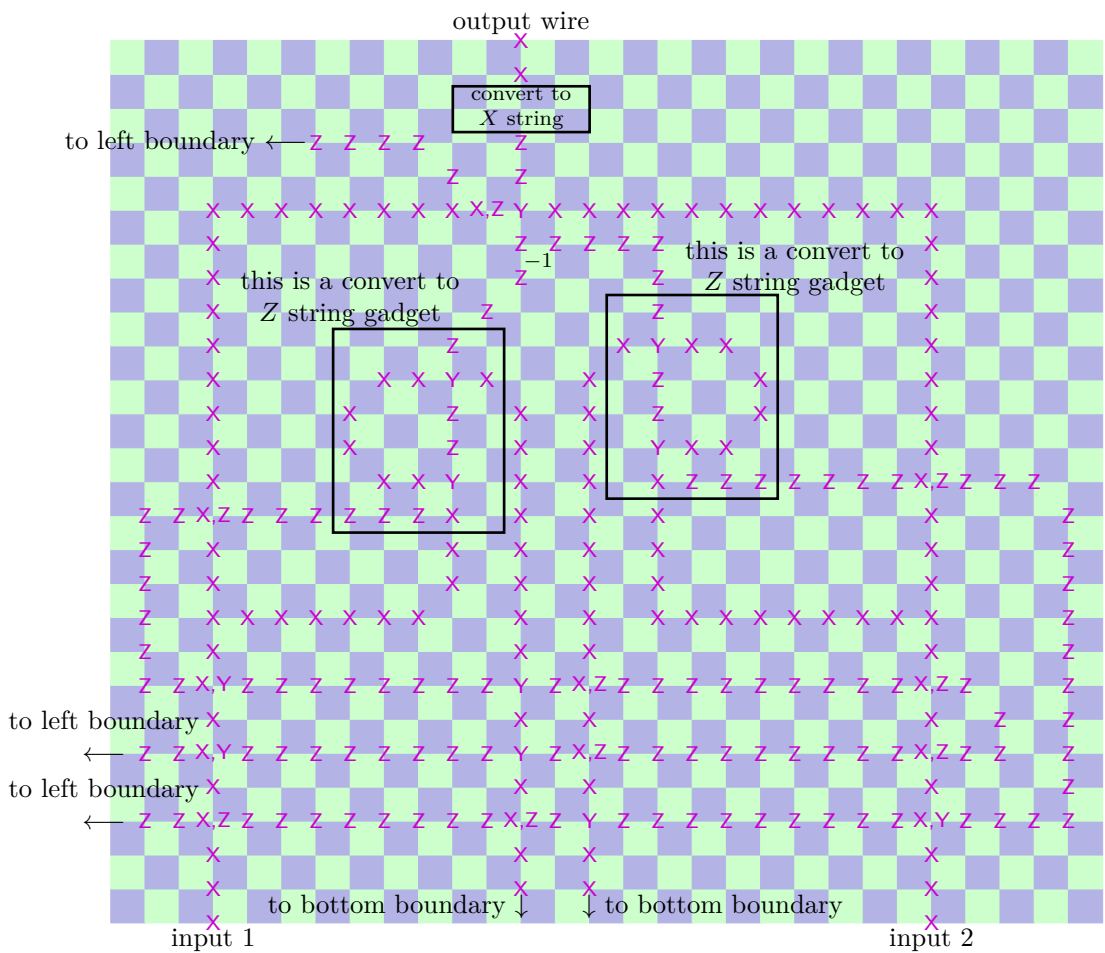
\begin{figure*}
			\centering\begin{tikzpicture}[x=0.45cm,y=0.45cm]
				\andGadget
			\end{tikzpicture}
			\caption{The AND gadget. The ``convert to $X$ string'' gadget results in an extra $X$ string coming out of the right of that gadget and an extra $Z$ string coming out of the left (both not shown for clarity), that will have to be routed to the appropriate boundaries.}\label{andGadget}
		\end{figure*}
	
		The AND gadget, given in \figref{andGadget}, is complicated, and we save the explanation for how it works for \appref{andGadgetAnalysis}. The 4 possible errors for this gadget (corresponding to the 4 possible values of the 2 input wires) are given in \figref{andGadgetPossibleErrors} in \appref{andGadgetAnalysis}.
	\subsection{Putting it all together: spacing between the gadgets}
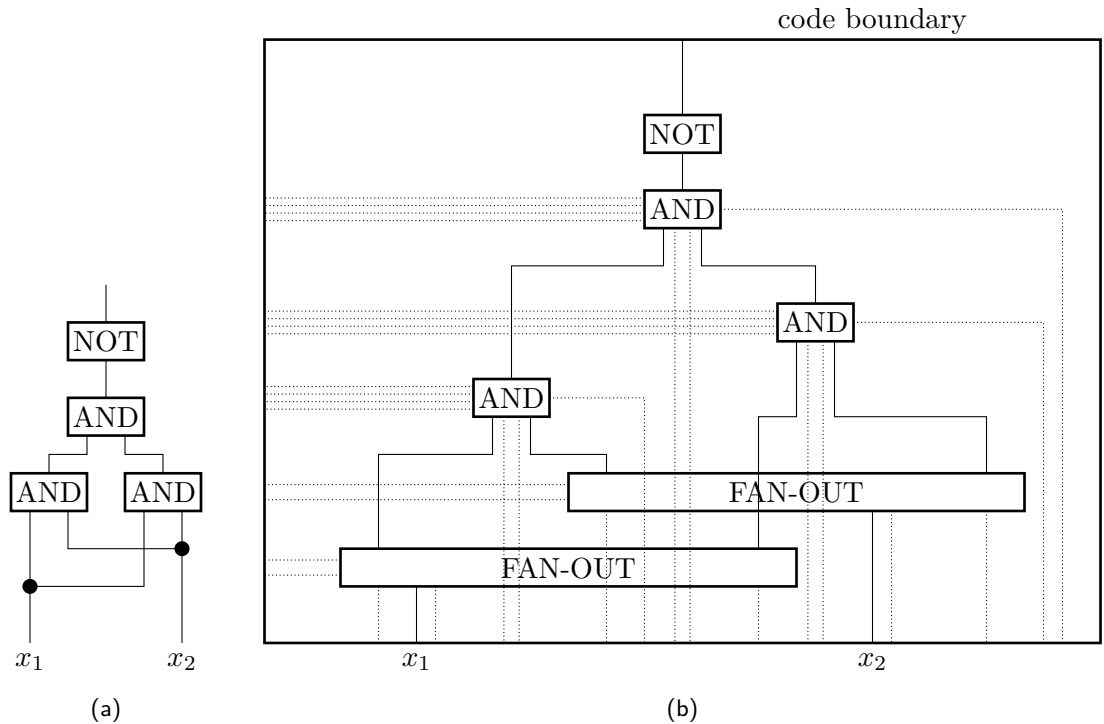
\begin{figure}
			\begin{subfigure}{0.2\textwidth}
				\centering\begin{tikzpicture}
					\node at (0, 0) {$x_1$};
					\draw (0, 0.25) -- (0, 1);
					\node at (2, 0) {$x_2$};
					\draw (2, 0.25) -- (2, 1.5);
					
					\node at (0, 1) [fill,circle,inner sep=2pt] {};
					\draw (0, 2) -- (0, 1) -- (1.5, 1) -- (1.5, 2);
					
					\node at (2, 1.5) [fill,circle,inner sep=2pt] {};
					\draw (0.5, 2) -- (0.5, 1.5) -- (2, 1.5) -- (2, 2);
					
					\draw[draw=black,line width=1pt] (-0.25,2) rectangle ++(1,0.5);
					\node at (0.25, 2.25) {AND};
					
					\draw[draw=black,line width=1pt] (1.25,2) rectangle ++(1,0.5);
					\node at (1.75, 2.25) {AND};
					
					\draw[draw=black,line width=1pt] (0.5,3) rectangle ++(1,0.5);
					\node at (1, 3.25) {AND};
					
					\draw[draw=black,line width=1pt] (0.5,4) rectangle ++(1,0.5);
					\node at (1, 4.25) {NOT};
					
					\draw (0.25, 2.5) -- (0.25, 2.75) -- (0.75, 2.75) -- (0.75, 3);
					\draw (1.75, 2.5) -- (1.75, 2.75) -- (1.25, 2.75) -- (1.25, 3);
					\draw (1, 3.5) -- (1, 4);
					\draw (1, 4.5) -- (1, 5);
				\end{tikzpicture}
				\caption{}\label{circuitUnstretched}
			\end{subfigure}
			\begin{subfigure}{0.8\textwidth}
				\centering\begin{tikzpicture}
					\tikzset{
						extraStringStyle/.style={densely dotted}
					}
				
					\draw[draw=black,line width=1pt] (0, 0) rectangle ++(11, 8);
					\node at (8, 8.25) {code boundary};
					
					\node at (2, -0.25) {$x_1$};
					\draw (2, 0) -- (2, 0.75);
					\node at (8, -0.25) {$x_2$};
					\draw (8, 0) -- (8, 1.75);
					
					\draw[draw=black,line width=1pt] (1, 0.75) rectangle ++(6,0.5);
					\node at (4, 1) {FAN-OUT};
					\draw[extraStringStyle] (2.25, 0.75) -- (2.25, 0);
					\foreach \y in {0.9, 1.1} \draw[extraStringStyle] (1, \y) -- (0, \y);
					\draw[extraStringStyle] (1.5, 0.75) -- (1.5, 0);
					\draw[extraStringStyle] (6.5, 0.75) -- (6.5, 0);
					
					\draw[draw=black,line width=1pt] (4, 1.75) rectangle ++(6,0.5);
					\node at (7, 2) {FAN-OUT};
					\draw[extraStringStyle] (8.25, 1.75) -- (8.25, 0);
					\foreach \y in {1.9, 2.1} \draw[extraStringStyle] (4, \y) -- (0, \y);
					\draw[extraStringStyle] (4.5, 1.75) -- (4.5, 0);
					\draw[extraStringStyle] (9.5, 1.75) -- (9.5, 0);
					
					\draw (1.5, 1.25) -- (1.5, 2.5) -- (3, 2.5) -- (3, 3);
					\draw (4.5, 2.25) -- (4.5, 2.5) -- (3.5, 2.5) -- (3.5, 3);
					\draw[draw=black,line width=1pt] (2.75, 3) rectangle ++(1,0.5);
					\node at (3.25, 3.25) {AND};
					\foreach \x in {3.15, 3.35} \draw[extraStringStyle]  (\x, 3) -- (\x, 0);
					\draw[extraStringStyle] (3.75, 3.25) -- (5, 3.25) -- (5, 0);
					\foreach \y in {3.1, 3.2, 3.3, 3.4} \draw[extraStringStyle] (2.75, \y) -- (0, \y);
					
					\draw (6.5, 1.25) -- (6.5, 3) -- (7, 3) -- (7, 4);
					\draw (9.5, 2.25) -- (9.5, 3) -- (7.5, 3) -- (7.5, 4);
					\draw[draw=black,line width=1pt] (6.75, 4) rectangle ++(1,0.5);
					\node at (7.25, 4.25) {AND};
					\foreach \x in {7.15, 7.35} \draw[extraStringStyle]  (\x, 4) -- (\x, 0);
					\draw[extraStringStyle] (7.75, 4.25) -- (10.25, 4.25) -- (10.25, 0);
					\foreach \y in {4.1, 4.2, 4.3, 4.4} \draw[extraStringStyle] (6.75, \y) -- (0, \y);
					
					\draw (3.25, 3.5) -- (3.25, 5) -- (5.25, 5) -- (5.25, 5.5);
					\draw (7.25, 4.5) -- (7.25, 5) -- (5.75, 5) -- (5.75, 5.5);
					\draw[draw=black,line width=1pt] (5, 5.5) rectangle ++(1,0.5);
					\node at (5.5, 5.75) {AND};
					\foreach \x in {5.4, 5.6} \draw[extraStringStyle]  (\x, 5.5) -- (\x, 0);
					\draw[extraStringStyle] (6, 5.75) -- (10.5, 5.75) -- (10.5, 0);
					\foreach \y in {5.6, 5.7, 5.8, 5.9} \draw[extraStringStyle] (5, \y) -- (0, \y);
					
					\draw[draw=black,line width=1pt] (5, 6.5) rectangle ++(1,0.5);
					\node at (5.5, 6.75) {NOT};
					
					\draw (5.5, 6) -- (5.5, 6.5);
					\draw (5.5, 7) -- (5.5, 8);
					
				\end{tikzpicture}
				\caption{}\label{circuitStretched}
			\end{subfigure}
			\caption{Our reduction stretches out the circuit horizontally and vertically in order to leave space for the $X$ and $Z$ strings to be routed to the appropriate boundaries. \subref{circuitUnstretched} is the original circuit, and \subref{circuitStretched} is the stretched version of the circuit embedded in the surface code outputted by the reduction, including the extra $X$ and $Z$ strings going to the boundaries. In the stretched version of the circuit \subref{circuitStretched}, solid lines are the $X$ wires between gadgets, and dashed lines are the extra $X$ and $Z$ strings routed to the boundaries. $X$ strings are those that go to the bottom boundary, and $Z$ strings are those that go to the left boundary. Recall that $X$ strings can cross the FAN-OUT gadgets, because the FAN-OUT gadgets (\figref{fanoutGadget}) consist of large horizontal stretches of $Z$ strings that can be crossed by $X$ strings via the $X/Z$ wire crossing gadget (\figref{xzCrossGadget}). This stretching increases the size of the output of the reduction by at most $O(m)$ in both the horizontal and vertical directions (where $m$ is the number of gates in the circuit), so at most the polynomial factor $O(m^2)$ in total.}\label{stretchReduction}
		\end{figure}

		When we stitch all these gadgets together, we need to leave enough space for all the relevant $X$ and $Z$ strings to cross each other and go to the boundary. Each gadget has a constant number of strings leaving it that go to the boundary, and the number of gadgets is given by some polynomial in the number of variables and clauses in the formula. Therefore we only need to add a polynomial amount of space between all the gadgets to leave space for all those strings to have space to cross each other and go to the boundary. Thus we can stitch together all these gadgets and get a surface code instance with syndromes and a noise model with total size that is bounded by some polynomial in the number of clauses and variables in the formula.  See \figref{stretchReduction} for an example of stretching the circuit to leave space for the extra $X$ and $Z$ strings.
		
		Once we stitch all the gadgets together, we have constructed a noise model and set of syndromes that satisfies properties \ref{propertyErrorsAssignments}, \ref{propertyTellSatisfying}, and \ref{propertyRigged} from Section \ref{overviewReductionSection}. Property \ref{propertyErrorsAssignments} is satisfied because the $n$ variable gadgets (each with their own binary choice of error present or not present) give $2^n$ possible errors, and all of the other gadgets in the circuit above the variable gadgets have their error present uniquely determined by their input, so those $2^n$ errors are the only possible errors. Property \ref{propertyTellSatisfying} is satisfied because we can tell whether an error corresponds to a satisfying or unsatisfying assignment by looking at whether $X$ errors are present in the output wire. Property \ref{propertyRigged} is satisfied because of the argument in Section \ref{overviewReductionSection} starting at \eqref{riggedProbabilities}. This completes the conversion from a boolean formula to a QMLD problem instance (box (a) in \figref{reductionSequence}).
	\subsection{Hardness of approximate QMLD}
		Our proof admits a straightforward generalization that establishes the hardness of approximate decoding. Here, \textbf{approximate QMLD} is the task of finding an error with probability at least $\frac{p'}{M}$, where $p'$ is the probability of the maximum probability error, and $M>1$ is the approximation factor. Here we consider $M$ to be a function of $\ell$, the number of qubits in the surface code instance, rather than $n$, to avoid confusion as we use $n$ to denote the number of variables in the boolean formula.
		\qmldApproxHardCorollary
		\begin{proof}
			For the sake of contradiction, assume we have a decoder that is guaranteed to output an error with probability at least $\frac{p'}{M(\ell)}$, where $p'$ is the probability of the maximum probability error, $M$ is some function always greater than 1 that grows at most exponentially fast (\ie $2^{\ell^{O(1)}}$), and $\ell$ is the number of qubits in the surface code instance. Then replace the probability $1-p^{\ell}$ in \figref{reductionOverview} with $1-\frac{p^{\ell}}{M(\ell)}$. This real number can be written down with a number of bits that is polynomial in $\ell$ (and thus polynomial in the size of the original boolean formula). Then the probability of any error corresponding to a satisfying assignment of the circuit is lower bounded by
			\begin{align*}
				\left(1-\frac{p^{\ell}}{M(\ell)}\right) p^{\ell-1}>p^{\ell}.
			\end{align*}
			And the probability of any error corresponding to an unsatisfying assignment of the circuit is upper bounded by $\frac{p^{\ell}}{M(\ell)}$. Therefore this approximate decoder will still find an error corresponding to a satisfying assignment of the circuit, if such an assignment exists.
		\end{proof}
	
		Note that it is common for decoders to optimize for $\log p'$, where $p'$ of the error they find, rather than to optimize for $p'$ itself. In this setting our hardness of approximation result for any exponential multiplicative approximation factor translates to a hardness of approximation result for any polynomial additive approximation factor.

\pagebreak
\section{Reduction from \#SAT to degenerate quantum maximum likelihood decoding (DQMLD)}
	In this section we prove our second main theorem.
	\dqmldHardTheorem
	
	\defbox{\dqmld*}
	
	We reduce from \#SAT, the canonical \#P-complete problem. \#SAT is the problem of, given a boolean formula with $n$ variables, determining how many assignments of true/false values to the input variables satisfy the formula (\ie make the formula output true), out of the $2^n$ possible assignments. This is more general than the SAT problem, which is merely determining whether that number of satisfying assignments is zero or nonzero. \#SAT (and its associated class of computational problems \#P) is generally thought of as much harder than SAT (and its associated class of computational problems NP).
	
	The strategy of the reduction is as follows. We start with the same noise model and syndromes as those outputted by the reduction in Section \ref{qmldHardSection}. We observe that all errors corresponding to satisfying assignments are in the same coset, as with all errors corresponding to unsatisfying assignments, which are in a different coset. Thus determining which coset has maximum coset probability already seems related to counting satisfying versus unsatisfying assignments. Given any integer $a\in\{0,1,\cdots,2^n\}$, we show how to modify the error probabilities in the noise model (but not which Pauli errors have nonzero probability) so that the maximum likelihood coset is the coset corresponding to satisfying assignments iff the formula has $\geq a$ satisfying assignments. Thus if we have a DQMLD algorithm, we can use that to determine whether the formula has $\geq a$ satisfying assignments for any integer $a$. This lets us exactly determine the number of satisfying assignments to any boolean formula using $O(n)$ calls to a DQMLD algorithm via binary search: we first try $a=\frac{1}{2}2^n$ to determine whether the formula has $\geq$ or $<\frac{1}{2}2^n$ satisfying assignments, then we try $a=\frac{3}{4}2^n$ in the former case or $a=\frac{1}{4}2^n$ in the latter case, and we continue this process of narrowing down the range of possible numbers of satisfying assignments until we exactly determine how many satisfying assignments the formula has.
	
	With this high-level strategy in mind, we can begin constructing the reduction from \#SAT to DQMLD for the surface code. First, we modify the error probabilities in the noise model outputted by the reduction from Section \ref{qmldHardSection} in the following way. Instead of giving $X$, $Y$, and $Z$ errors probability 0 or $p$, we give them the following probabilities:
	\begin{itemize}
		\item If only one error is possible for a qubit (\eg if we only draw $X$ on a qubit), then that error occurs with probability $\frac{1}{2}$ and no error occurs with probability $\frac{1}{2}$. This excludes the topmost possible $X$ error which is special (we set that probability later).
		\item If two different errors are possible (\eg if we draw $X,Y$ on a qubit), then each of those errors occurs with probability $\frac{1}{3}$, and no error occurs with probability $\frac{1}{3}$.
		\item If all three errors are possible on a qubit (\ie if we draw $X,Y,Z$ on a qubit), then each of those errors occurs with probability $\frac{1}{4}$, and no error occurs with probability $\frac{1}{4}$.
		\item The topmost possible $X$ error on the output wire occurs with probability $r$ (to be determined later), instead of probability $1-p^{\ell}$.
	\end{itemize}
	
	The important consequence of setting the error probabilities this way is that all errors corresponding to satisfying assignments of the circuit have the same probability of occurring, and likewise with all errors corresponding to unsatisfying assignments. The probability of any error occurring, ignoring the special topmost qubit of the output wire, is
	\begin{align*}
		&1^{\text{\# of qubits where no error is possible}} \\
		\times &\left(\frac{1}{2}\right)^{\text{\# of qubits where 1 error is possible, except topmost qubit of output wire}} \\
		\times &\left(\frac{1}{3}\right)^{\text{\# of qubits where 2 errors are possible}} \\
		\times &\left(\frac{1}{4}\right)^{\text{\# of qubits where 3 errors are possible}}.
	\end{align*}
	Call this probability $q$. Now taking into account the special topmost qubit of the output wire, all errors corresponding to satisfying assignments have probability $qr$. All errors corresponding to unsatisfying assignments have probability $q(1-r)$.
	
	The next ingredient we need in the \#P-hardness proof is the following lemma, which we prove in \appref{lemmaProof}.
	
	\begin{restatable}{lemma}{dqmldLemma}
		\label{dqmldLemma}
		All errors corresponding to satisfying assignments of the circuit are equivalent up to stabilizers, and are thus in the same coset $C$. All errors corresponding to unsatisfying assignments of the circuit are in the coset $\lx C$, where $\lx$ is a logical $X$ operator.
	\end{restatable}

	Using this lemma, we can now prove Theorem \ref{dqmldHardTheorem}.
	\begin{proof}
		Let the number of satisfying assignments of the boolean formula be $a$, and let the number of unsatisfying assignments of the formula be $b$. Note that $b=2^n-a$, where $n$ is the number of variables in the formula. Lemma \ref{dqmldLemma} tells us that the coset of errors with maximum likelihood is $C$ iff $qra> q(1-r)b$.\footnote{The maximum likelihood coset is undefined if $qra=q(1-r)b$, because both cosets have the same coset probability in this case. This minor technical issue can be avoided by setting the pivot points in the binary search not to $\frac{a}{2^n}$ for any integer $a$, but rather $\frac{a}{2^n}+\frac{1}{2^{n+1}}$ for any integer $a$.} This is equivalent to
		\begin{align*}
			\frac{r}{1-r}&>\frac{b}{a} \\
			\iff \frac{r}{1-r}&>\frac{b}{2^n-b} \\
			\iff \frac{r}{1-r}&>\frac{\frac{b}{2^n}}{1-\frac{b}{2^n}}.
		\end{align*}
		Since $\frac{r}{1-r}$ is a strictly increasing function of $r$ when $r\in(0,1)$, this condition is equivalent to $r>\frac{b}{2^n}$. Thus by setting $r$ equal to any real number of our choice in $(0,1)$ (which we only need to specify to $n+1$ bits of precision), solving DQMLD, and seeing whether the error returned corresponds to a satisfying or unsatisfying assignment of the formula, we can determine whether the proportion of assignments that don't satisfy the formula is $<r$.
		
		Thus if we have a DQMLD algorithm for the surface code, we can exactly determine the number of satisfying assignments for any boolean formula by doing $O(n)$ rounds of binary search on $r$, where $n$ is the number of variables in the formula. Thus DQMLD for the surface code is \#P-hard.
	\end{proof}
	
	This reduction involves multiple calls to an oracle for DQMLD, rather than outputting a single instance of a DQMLD problem where the answer to that DQMLD problem encodes the answer to another \#P-hard problem. Therefore this is a Turing reduction, not a many-one/Karp reduction. As there are multiple oracle calls in this reduction, the Turing nature of the reduction seems essential, unlike with the QMLD reduction where it could be transformed into a many-one/Karp reduction by formulating QMLD as a decision problem. Turing reductions are common for \#P-hardness proofs. For example, both Leslie Valiant's original proof that computing the permanent is \#P-hard\cite{valiant1979permanent}, and Iyer and Poulin's result that DQMLD for general stabilizer codes is \#P-hard\cite{hardnessStabilizerCodesIyerPoulin15}, used Turing reductions.
	
	\subsection{Hardness of approximate DQMLD}
		As with QMLD, we can strengthen this result to show that even approximate DQMLD for the surface code is hard. Here, we use the fact that \#SAT is hard to approximate to within any nontrivial exponential factor. Specifically, for boolean formulas with $n$ variables, and for all $\epsilon>0$, it is NP-hard to approximate the number of solutions to such formulas with approximation ratio $2^{n^{1-\epsilon}}$\cite{hardApproxCounting}.
		
		This hardness of approximation result for \#SAT translates directly into a hardness of approximation result for DQMLD for the surface code. Here, \textbf{approximate DQMLD} is the task of finding an error in a coset $C$ such that the coset probability of $C$ is at least $\frac{p'}{M}$, where $p'$ is the coset probability of the maximum likelihood coset and $M>1$ is the approximation factor. As in the previous section, here we consider $M$ to be a function of $\ell$, the number of qubits in the surface code instance, rather than $n$, to avoid confusion as we use $n$ to denote the number of variables in the boolean formula.
		\dqmldApproxHardCorollary
		\begin{proof}
			For the sake of contradiction, assume we have a decoder that is guaranteed to output an error in a coset $C$ with coset probability at least $\frac{p'}{M(\ell)}$, where $p'$ is the coset probability of the maximum likelihood coset. Here $M(\ell)>1$ is the approximation factor as a function of $\ell$, the number of qubits in the surface code instance. If we use this decoder to perform binary search on the proportion of unsatisfying assignments as we did in the proof of Theorem \ref{dqmldHardTheorem}, then this binary search will settle on a value that is correct up to a factor $O\left(M(\ell)\right)$.
			
			This means approximate DQMLD for the surface code is NP-hard if $M(\ell)=O\left(2^{n^{1-\epsilon}}\right)$ for some $\epsilon>0$. Since $\ell$ is $n^{O(1)}$, or equivalently $n$ is $\ell^{{O(1)}}$ (since we can assume we are dealing with boolean formulas with total size upper bounded by some polynomial in the number of variables), this means there is some constant $c$ such that approximate DQMLD for the surface code is NP-hard if the approximation ratio $M(\ell)$ is O$\left(2^{\ell^c}\right)$.
		\end{proof}
\section{Hardness of decoding with more regular noise models} \label{regNoiseSection}
	The noise models outputted by our reduction are contrived and unphysical, because they result in many qubits having zero error probability. This raises the question, what can we say about the hardness of decoding the surface code with more natural, physically realistic noise models? Here, ``natural'' could mean the error probabilities don't vary too much between qubits, or the error probabilities are never too small or large. In this section we make a small step towards hardness results with more natural noise models by doing away with the ``special'' top qubit of the output wire that has different error probabilities than the rest of the qubits.
	
	\subsection{More regular noise models for QMLD reduction}
		\begin{figure}
			\centering\begin{tikzpicture}
				
				\draw[draw=black,line width=1pt] (0,0) rectangle ++(5,6);
				
				\draw[draw=black,line width=1pt] (0.1,0.1) rectangle ++(4.8,2.8);
				
				\draw (1, 0) -- (1, 1);
				\draw (2, 0) -- (2, 1);
				\node at (3, 0.5) {$\cdots$};
				\draw (4, 0) -- (4, 1);
				\draw[draw=black,line width=1pt] (0.5,1) rectangle ++(4,1);
				\node at (2.5, 1.5) {circuit};
				\draw (2.5, 2) -- (2.5, 3);
				
				\node at (3.5, 2.5) {$\longleftarrow$ output};
				
				\node[align=center] at (-2, 4) {original QMLD \\ reduction; \\ uses $\ell$ qubits};
				\draw [->] (-1, 4) -- (1, 3);

				\draw (2.5, 3) -- (2.5, 3);
				\node at (2.5, 3.5) {NOT};
				\draw[draw=black,line width=1pt] (2, 3) rectangle ++(1,1);
				\draw (2.5, 4) -- (2.5, 6);
				
				\node[align=center] at (1, 5) {length $2\ell$ \\ wire};
				\draw (2.3, 4.1) -- (2.1, 4.1) -- (2.1, 5.9) -- (2.3, 5.9);
				\draw (2.1, 5) -- (1.8, 5);
			\end{tikzpicture}
			\caption{A modification of the reduction (as outlined in \figref{reductionOverview}) that lets all qubits have the same probabilities for possible errors (0 or $p$), instead of having one special qubit with error probability $1-p^{\ell}$.}\label{homogeneousQMLD}
		\end{figure}
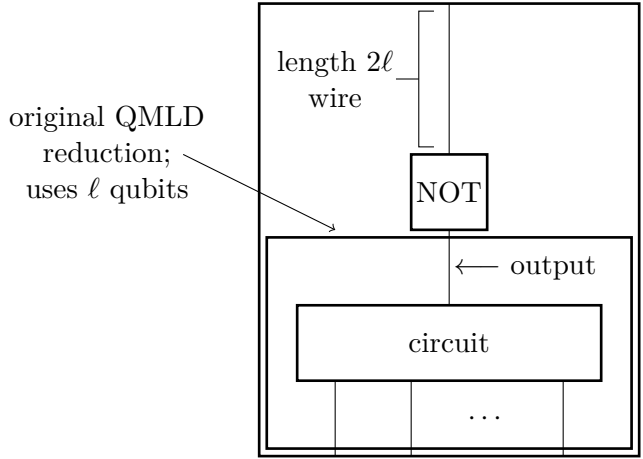
		
		We can modify the QMLD reduction so that the topmost qubit of the output wire doesn't have the physically unrealistic error rate $1-p^{\ell}$. This modification is in \figref{homogeneousQMLD}. The modification adds a NOT gate on the output wire and a length $2\ell$ wire onto the original output wire, where $\ell$ is the number of qubits used in the unmodified version of the reduction as outlined in \figref{reductionOverview}. This modification means that the probability of any error corresponding to a satisfying assignment picks up a $(1-p)^{2\ell}$ term, and the probability of any error corresponding to an unsatisfying assignment picks up a $p^{2\ell}$ term.
		
		This lets us lower bound the probability of any error corresponding to a satisfying assignment by
		\begin{align*}
			\pr{\text{error corresp.~to satisfying assignment}}\geq p^{\ell}(1-p)^{2\ell}
			=p^{\ell}\left((1-p)^2\right)^{\ell}
			> p^{\ell}p^{\ell}
			=p^{2\ell}.
		\end{align*}
		Here we used the fact that $p\in(0,0.25]\implies (1-p)^2>p$. The probability of any error corresponding to an unsatisfying assignment is trivially upper bounded by $p^{2\ell}$. This means that any error corresponding to a satisfying assignment will have a higher probability than any error corresponding to an unsatisfying assignment. Thus QMLD in this setting remains NP-hard.
		
		We can maintain the hardness of approximation result for any exponential approximation factor $M(\ell)$ by giving that wire length $2\ell+2\log_p\frac{1}{M(\ell)}$, instead of length $\ell$. Then the probability of any error corresponding to an unsatisfying assignment is upper bounded by
		\begin{align*}
			p^{2\ell+2\log_p\frac{1}{M(\ell)}}
			=p^{2\ell}\frac{1}{M(\ell)^2}.
		\end{align*}
		And the probability of any error corresponding to a satisfying assignment is lower bounded by
		\begin{align*}
			p^{\ell}(1-p)^{2\ell+2\log_p\frac{1}{M(\ell)}}
			>p^{\ell}p^{\ell+\log_p\frac{1}{M(\ell)}}
			=p^{2\ell}\frac{1}{M(\ell)}.
		\end{align*}
		Therefore a $M(\ell)$ approximate QMLD algorithm will still always find an error corresponding to a satisfying assignment, if one exists.
	\subsection{More regular noise models for DQMLD reduction}
		In the DQMLD reduction, the topmost qubit of the output wire is ``special'', as the probability of that $X$ error takes on special values. If we don't make that qubit special and instead give it an error probability of $\frac{1}{2}$, like all other qubits with one possible error, then DQMLD decoding applied to the resulting noise model can be used to decide Majority-SAT, the problem of deciding whether a boolean formula has more satisfying or unsatisfying assignments. Since an oracle for Majority-SAT can be used to solve \#SAT (\cite{AroraBarak} Lemma 17.7 proves this), surface code decoding is still \#P-hard even if we restrict all qubits to have error probabilities 0, $\frac{1}{2}$, $\frac{1}{3}$, or $\frac{1}{4}$ and don't have a special qubit with more specific error probabilities. However, we then lose the hardness of approximation result Corollary \ref{dqmldApproxHardCorollary}. This is because an approximate DQMLD algorithm with constant approximation ratio $M$ will only be able to solve the BPP-complete promise problem of deciding whether a boolean formula has at least $2^n\frac{M}{M+1}$ satisfying assignments or at most $2^n\frac{1}{M+1}$ satisfying assignments, promised that one of those is the case.
	
	Although these minor modifications to the reductions do make the resulting noise models slightly more homogeneous, they still result in many qubits in the reduction having 0 error probability, which is clearly physically unrealistic. We leave it as an open problem whether one can show any hardness results for decoding with more physically realistic noise models.

\section{Conclusion and discussion}
	We have shown computational hardness results for maximum probability error decoding and maximum likelihood decoding of the surface code, and for approximate versions of those problems. Therefore no efficient surface code decoding algorithm can always solve, or even always approximate, QMLD or DQMLD for all possible Pauli noise models and syndromes (modulo the standard computational complexity assumptions $\p\neq\np$ and $\fp\neq\sharpP$). This provides some explanation as to why all known surface code decoding algorithms are not known to be optimal or approximately optimal except in a few special cases.
	
	These no-go results are highly relevant for quantum computing, because the surface code is one of the most promising candidates for an error correcting code with which to do fault-tolerant quantum computation, and we need fast and accurate decoders in order to use the surface code. However, these results are not really bad news for quantum computing, because we can still achieve fault tolerance with the imperfect decoders we have now. This is because the current decoders get the right answer almost all of the time---\ie for almost all sets of syndromes with physically reasonable noise models. Instead the practical consequence of these results, like all hardness results for practical problems of interest, is to inform research into surface code decoding algorithms. We now know that one cannot hope for an algorithm that always solves the DQMLD or DQMLD problem (or even always approximates it) for all possible Pauli noise models and all possible syndromes. Rather, one should look for algorithms that take advantage of special properties of the noise or syndromes (independence of $X$ and $Z$ errors is one such special property that some decoders take advantage of), or look for heuristic algorithms without rigorous performance guarantees (tensor network decoders\cite{BSV14} and belief-matching\cite{fragileBoundariesSurfaceCodeBeliefMatching} are two examples).
	
	A clear open problem suggested by our work is to establish hardness results for surface code decoding with more physically realistic noise models. In the context of proving hardness results, restricting the problem instances considered to a subset of all problem instances makes the hardness results stronger. So one could strengthen our results by proving similar hardness results for a restricted class of more physically realistic noise models, such as depolarizing noise, Pauli noise where all error probabilities are sufficiently far from 0, or circuit noise where all components of the syndrome extraction circuits have nonzero error rates. Even for simple depolarizing noise there are no known optimal decoders. Can we get any hardness results for noise models as simple as depolarizing noise? Or can we further characterize the special cases where we can get provably optimal (or provably approximately optimal) decoders for the surface code?
	
\section{Acknowledgments}
	This work is supported by a collaboration between the U.S. DOE and the National Science Foundation. The material is based upon work supported by the U.S. Department of Energy, Office of Science, National Quantum Information Science Research Centers, Quantum Systems Accelerator. It is also supported by the NSF STAQ Project (PHY-1818914, PHY-2325080).
	
	Alex Fischer would like to thank Milad Marvian for teaching an excellent class on quantum error correction which got him interested in the subject.

\bibliographystyle{quantum}
\bibliography{./bib/bib}{}

\pagebreak

\begin{appendices}

\section{Case analysis of the AND gadget}\label{andGadgetAnalysis}
	Here we exhaustively analyze the AND gadget to show that it functions correctly as an AND gadget: that is, that the only 4 possible errors in this gadget correspond to the 4 possible settings of the 2 input wires along with the correct setting of the output wire. These 4 errors are given in \figref{andGadgetPossibleErrors} (split across 4 pages).
	
	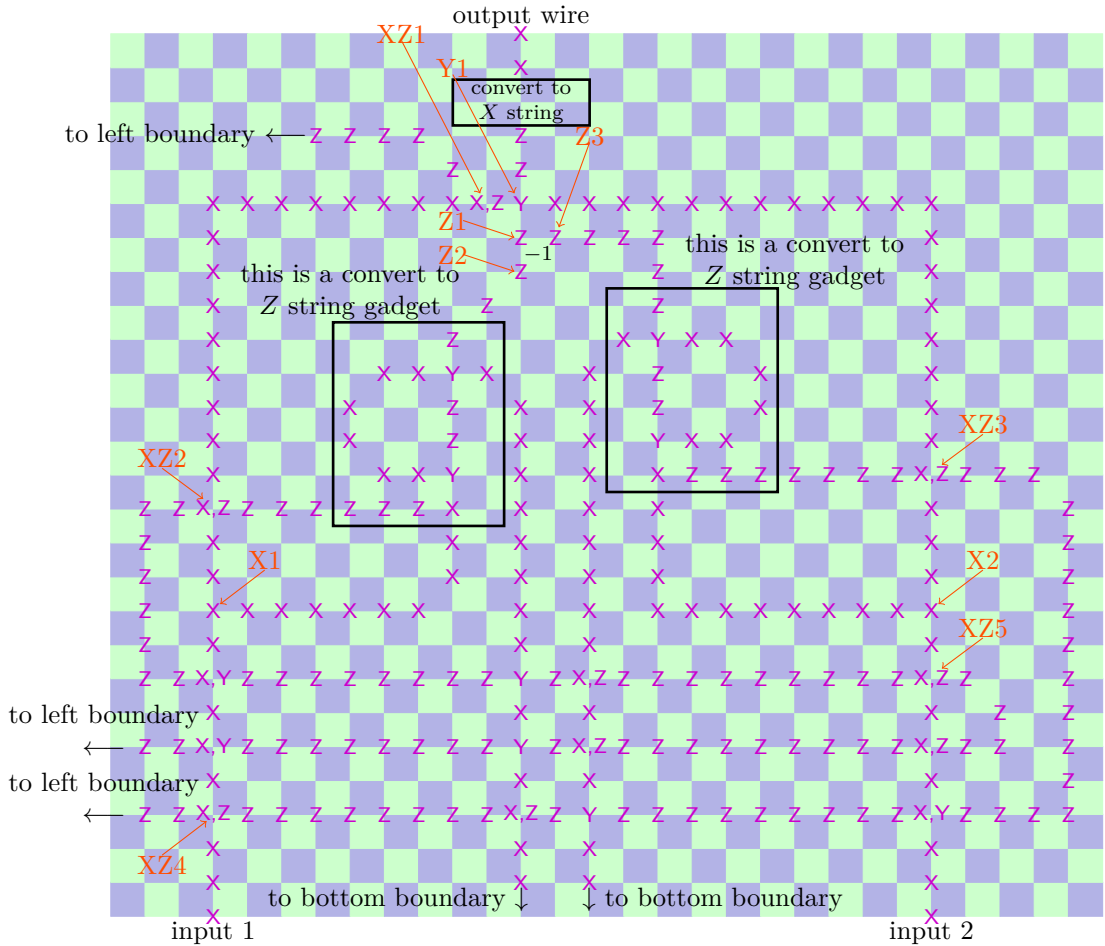
\begin{figure*}
		\centering\begin{tikzpicture}[x=0.45cm,y=0.45cm]
			\andGadget
			
			\node [locationLabelStyle] at (10, 25) {Y1};
			\draw [->,color=locationLabelColor] (10, 24.8) -- (11.8, 21.3);
			
			\node [locationLabelStyle] at (8.5, 26) {XZ1};
			\draw [->,color=locationLabelColor] (8.5, 25.8) -- (10.8, 21.3);
			
			\node [locationLabelStyle] at (10, 20.5) {Z1};
			\draw [->,color=locationLabelColor] (10.3, 20.5) -- (11.8, 20);
			
			\node [locationLabelStyle] at (10, 19.5) {Z2};
			\draw [->,color=locationLabelColor] (10.3, 19.5) -- (11.8, 19);
			
			\node [locationLabelStyle] at (14, 23) {Z3};
			\draw [->,color=locationLabelColor] (14, 22.8) -- (13.1, 20.3);
			
			\node [locationLabelStyle] at (1.5, 13.5) {XZ2};
			\draw [->,color=locationLabelColor] (1.5, 13.2) -- (2.7, 12.3);
			
			\node [locationLabelStyle] at (25.5, 14.5) {XZ3};
			\draw [->,color=locationLabelColor] (25.5, 14.2) -- (24.3, 13.3);
			
			\node [locationLabelStyle] at (1.5, 1.5) {XZ4};
			\draw [->,color=locationLabelColor] (1.5, 1.8) -- (2.8, 2.8);
			
			\node [locationLabelStyle] at (25.5, 8.5) {XZ5};
			\draw [->,color=locationLabelColor] (25.5, 8.2) -- (24.3, 7.3);
			
			\node [locationLabelStyle] at (4.5, 10.5) {X1};
			\draw [->,color=locationLabelColor] (4.5, 10.2) -- (3.2, 9.2);
			
			\node [locationLabelStyle] at (25.5, 10.5) {X2};
			\draw [->,color=locationLabelColor] (25.5, 10.2) -- (24.2, 9.2);
		\end{tikzpicture}
		\caption{The AND gadget with important locations labeled that are used in the case analysis.}\label{andGadgetLabeled}
	\end{figure*}
		
	In this case analysis we use labels given to important parts of the gadget in \figref{andGadgetLabeled}. The $-1$ syndrome in the gadget must have 1 or 3 $Z$ errors overlapping with it. We consider all 4 cases (1 of the 3 possible $Z$ errors, all 3 possible $Z$ errors) separately.
	\subsection{Case 1: all 3 locations have $Z$ errors}
		This is the case where locations Z1, Z2, and Z3 all have $Z$ errors. Z2 has a $Z$ error implies location XZ2 has a $Z$ error and location X1 has an $X$ error. The string of $X$ errors can either go up or down from location X1; going up is impossible since XZ2 has a $Z$ error, not an $X$ error, so this string must go downwards to input 1, which means location XZ4 has an $X$ error.
		
		Z3 has a $Z$ error implies XZ3 has a $Z$ error. Follow this string of possible $Z$ errors and you see that this implies that location XZ4 has a $Z$ error. But we already showed that XZ4 has to have an $X$ error. That means \textbf{this case is impossible}.
		
		Note that location XZ4 can have an $X$ error, or a $Z$ error, but not a $Y$ error. This is an example of explicitly non-independent $X$ and $Z$ error probabilities, which are an essential ingredient in this reduction. It also means the known decoders do not always perform optimally on these types of error models.
	\subsection{Case 2: of those 3 possible $Z$ errors, only location Z1 has a $Z$ error}
		In order to get a $+1$ measurement result for the $X$ stabilizer that is above and left of the stabilizer that was measured to be $-1$, we need either location Y1 to have a $Y$ error, or XZ1 to have a $Z$ error (but not both). We consider these 2 sub-cases separately.
	\subsection{Case 2a: location Z1 has a $Z$ error, and XZ1 has a $Z$ error}
		The string of $Z$ errors that starts at Z1 propagates up and left through location XZ1, to the $Z$ boundary. Location Y1 cannot have a $Y$ error, else the stabilizer up and left of the stabilizer that was measured to be $-1$ would not have been measured $+1$. This means there cannot be a string of $Z$ errors coming out above location Y1, which means the output wire is false.
		
		We claim both the input wires have to be false. We prove this by contradiction, for each wire.
		
		If input 1 is true, then location X1 has an $X$ error. That error string can propagate up or right from location X1. If that error string propagates right, then location Z2 has $Z$ error---a contradiction. If that error string propagates up, then location XZ1 has an $X$ error---a contradiction. Thus input 1 is false.
		
		If input 2 is true, then location X2 has an $X$ error. That error string can propagate up or left from location X2. If that error string propagates left, then location Z3 has a $Z$ error---a contradiction. If that error string propagates up, then that string propagates to location Y1 which means that location Y1 has a $Y$ error---a contradiction. Thus input 2 is false.
		
		Thus in this case 2a, the only satisfying assignment is an assignment that has input 1 false, input 2 false, and the output wire false.
	\subsection{Case 2b: location Z1 has a $Z$ error, and location Y1 has a $Y$ error}
		Location Y1 having a $Y$ error means there has to be a $Z$ string coming up above location Y1, which means that the output wire is true. It also means that there needs to be $X$ strings coming out of the left and right side of location Y1.
		
		The $X$ error string coming left out of location Y1 will propagate through location XZ2 (which means that location XZ2 has an $X$ error) to location X1. From location X1, that error string can propagate downwards or to the right. If that error string propagates rightwards from location X1, it means location XZ2 will be a $Z$ error---which is a contradiction. Thus the $X$ error string will propagate downwards to input 1, which means input 1 has to be true.
		
		We can apply mirror image of that argument to the $X$ error string coming out of the right side of location Y1, which tells us that input 2 has to be true.
		
		Thus in this case 2b, the only satisfying assignment has input 1 true, input 2 true, and the output wire true.
	\subsection{Case 3: of those 3 possible $Z$ errors, only location Z2 has a $Z$ error}		
		The $Z$ string propagating down and left from location Z2 means there needs to be a $Z$ string propagating through location XZ2, all the way to location XZ5, then to the $Z$ boundary. In particular both locations XZ2 and XZ5 have $Z$ errors. It also means there needs to be an $X$ string propagating down and left to location X1. From there, that error string can propagate up or down. If it propagates up, it means location XZ2 has has an $X$ error, which is a contradiction. Thus the error string propagates down from there, which means the input 1 wire has $X$ errors and is true.
		
		We claim the input 2 wire has to not have $X$ errors and thus be false. If it did have $X$ errors, location XZ5 would have an $X$ error---which is a contradiction.
		
		We claim the output wire has to not have $X$ errors and thus be false. If it did have $X$ errors, then location Y1 would have a $Y$ error, which means there would be a string of $X$ errors propagating left from location Y1. This means location XZ2 would have an $X$ error---a contradiction.
		
		Thus in this case 3, the only satisfying assignment has input 1 true, input 2 false, and the output wire false.
	\subsection{Case 4: of those 3 possible $Z$ errors, only location Z3 has a $Z$ error}
		In this case the same argument as the above case 3 applies, but the mirror image. Thus the only satisfying assignment has input 1 false, input 2 true, and the output wire false.
		
		This completes the case analysis. Through exhaustive case analysis, we have seen that the only possible errors that are consistent with that $-1$ syndrome are the 4 assignments that have the correct input 1, input 2, and output truth values for this gadget to function as an AND gadget.
	
	\begin{figure*}
		\begin{subfigure}{\textwidth}
			\centering\begin{tikzpicture}[x=0.45cm,y=0.45cm]z
				\andGadget
				\node at (12, 20) [errorStyle] {Z};
				\node at (11, 21) [errorStyle] {Z};
				\node at (10, 22) [errorStyle] {Z};
				\foreach \x in {9,...,6} \node at (\x, 23) [errorStyle] {Z};
			\end{tikzpicture}
			\caption{The error on the AND gadget when both inputs are false. Recall that purple operators drawn on qubits denote the noise model, whereas red operators drawn on qubits denote the actual error that happened.}\label{andGadgetFF}
		\end{subfigure}
	\end{figure*}\begin{figure*}\ContinuedFloat
		\begin{subfigure}{\textwidth}
			\centering\begin{tikzpicture}[x=0.45cm,y=0.45cm]
				\andGadget
				\foreach \y in {0,...,4} \node at (3, \y) [errorStyle] {X};
				\node at (3, 5) [errorStyle] {Y};
				\node at (3, 6) [errorStyle] {X};
				\node at (3, 7) [errorStyle] {Y};
				\node at (3, 8) [errorStyle] {X};
				
				\foreach \x in {3,...,9} \node at (\x, 9) [errorStyle] {X};
				\foreach \y in {10,11,12} \node at (10, \y) [errorStyle] {X};
				\node at (10, 13) [errorStyle] {Y};
				\node at (10, 14) [errorStyle] {Z};
				\node at (10, 15) [errorStyle] {Z};
				\node at (10, 16) [errorStyle] {Y};
				\node at (10, 17) [errorStyle] {Z};
				\node at (11, 18) [errorStyle] {Z};
				\node at (12, 19) [errorStyle] {Z};
				
				\foreach \x in {9,8} \node at (\x, 13) [errorStyle] {X};
				\foreach \y in {14,15} \node at (7, \y) [errorStyle] {X};
				\foreach \x in {8,9} \node at (\x, 16) [errorStyle] {X};
				\node at (11, 16) [errorStyle] {X};
				\foreach \y in {15,...,8} \node at (12, \y) [errorStyle] {X};
				\node at (12, 7) [errorStyle] {Y};
				\node at (12, 6) [errorStyle] {X};
				\node at (12, 5) [errorStyle] {Y};
				\foreach \y in {4,...,1} \node at (12, \y) [errorStyle] {X};
				
				\foreach \x in {9,...,1} \node at (\x, 12) [errorStyle] {Z};
				\foreach \y in {11,...,7} \node at (1, \y) [errorStyle] {Z};
				\node at (2, 7) [errorStyle] {Z};
				\foreach \x in {4,...,11} \node at (\x, 7) [errorStyle] {Z};
				\foreach \x in {13,...,25} \node at (\x, 7) [errorStyle] {Z};
				\node at (26, 6) [errorStyle] {Z};
				\foreach \x in {26,...,13} \node at (\x, 5) [errorStyle] {Z};
				\foreach \x in {11,...,4} \node at (\x, 5) [errorStyle] {Z};
				\foreach \x in {2,1} \node at (\x, 5) [errorStyle] {Z};
				
			\end{tikzpicture}
			\caption{The error on the AND gadget when input 1 is true and input 2 is false.}\label{andGadgetTF}
		\end{subfigure}
	\end{figure*}\begin{figure*}\ContinuedFloat
		\begin{subfigure}{\textwidth}
			\centering\begin{tikzpicture}[x=0.45cm,y=0.45cm]
				\andGadget
				\foreach \y in {0,1,2} \node at (24, \y) [errorStyle] {X};
				\node at (24, 3) [errorStyle] {Y};
				\foreach \y in {4,...,8} \node at (24, \y) [errorStyle] {X};
				
				\foreach \x in {24,...,16} \node at (\x, 9) [errorStyle] {X};
				\foreach \y in {10,...,13} \node at (16, \y) [errorStyle] {X};
				\node at (16, 14) [errorStyle] {Y};
				\node at (16, 15) [errorStyle] {Z};
				\node at (16, 16) [errorStyle] {Z};
				\node at (16, 17) [errorStyle] {Y};
				\foreach \y in {18,19,20} \node at (16, \y) [errorStyle] {Z};
				\foreach \x in {15,14,13} \node at (\x, 20) [errorStyle] {Z};
				
				\foreach \x in {17,18} \node at (\x, 14) [errorStyle] {X};
				\foreach \y in {15,16} \node at (19, \y) [errorStyle] {X};
				\foreach \x in {18,17} \node at (\x, 17) [errorStyle] {X};
				\node at (15, 17) [errorStyle] {X};
				\foreach \y in {16,...,4} \node at (14, \y) [errorStyle] {X};
				\node at (14, 3) [errorStyle] {Y};
				\foreach \y in {2,1} \node at (14, \y) [errorStyle] {X};
				
				\foreach \x in {17,...,27} \node at (\x, 13) [errorStyle] {Z};
				\foreach \y in {12,...,3} \node at (28, \y) [errorStyle] {Z};
				\foreach \x in {27,26,25} \node at (\x, 3) [errorStyle] {Z};
				\foreach \x in {23,...,15} \node at (\x, 3) [errorStyle] {Z};
				\foreach \x in {13,...,1} \node at (\x, 3) [errorStyle] {Z};
				
			\end{tikzpicture}
			\caption{The error on the AND gadget when input 1 is false and input 2 is true.}\label{andGadgetFT}
		\end{subfigure}
	\end{figure*}\begin{figure*}\ContinuedFloat
		\begin{subfigure}{\textwidth}
			\centering\begin{tikzpicture}[x=0.45cm,y=0.45cm]
				\andGadget
				\foreach \y in {0,...,21} \node at (3, \y) [errorStyle] {X};
				\foreach \x in {4,...,11} \node at (\x, 21) [errorStyle] {X};
				
				\foreach \y in {0,...,21} \node at (24, \y) [errorStyle] {X};
				\foreach \x in {23,...,13} \node at (\x, 21) [errorStyle] {X};
				
				\node at (12, 20) [errorStyle] {Z};
				\node at (12, 21) [errorStyle] {Y};
				\foreach \y in {22,23} \node at (12, \y) [errorStyle] {Z};
				\foreach \y in {25,26} \node at (12, \y) [errorStyle] {X};
				
			\end{tikzpicture}
			\caption{The error on the AND gadget when both inputs are true.}\label{andGadgetTT}
		\end{subfigure}
		\caption{The 4 possible errors in the AND gadget. The case analysis proves that these are the only 4 possible errors.}\label{andGadgetPossibleErrors}
	\end{figure*}
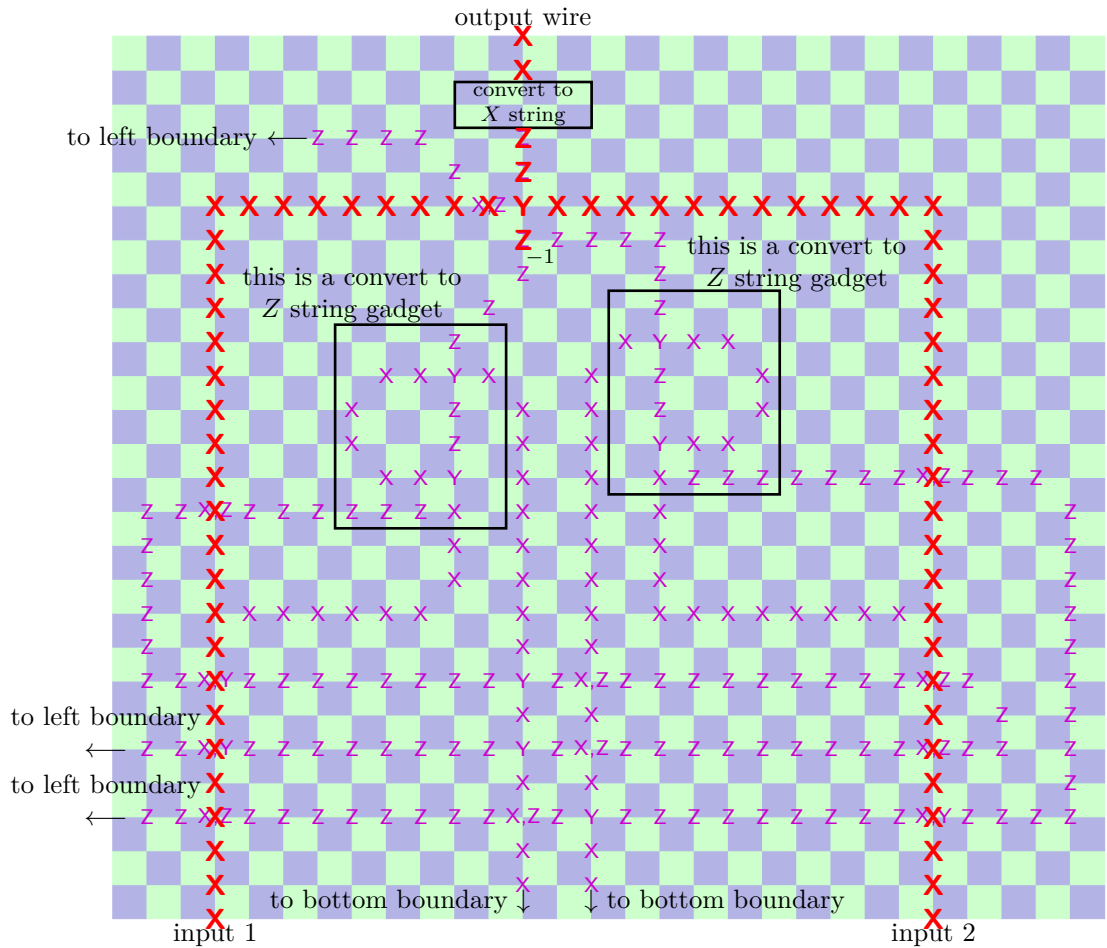

	\pagebreak
\section{Proof of Lemma \ref{dqmldLemma}}\label{lemmaProof}
	\dqmldLemma*
	
	We prove this by looking at the change in the error when we change the value of one of the input variables. We show:
	\begin{itemize}
		\item If changing that variable's value does not change the value of the output of the circuit, then the change in the error is a combination of strings of $X$ errors that start and end at the bottom boundary (\ie a product of $X$ stabilizers) and strings of $Z$ errors that start and end at the left boundary (\ie a product of $Z$ stabilizers).
		\item If changing that variable's value does change the output of the circuit, then the change in the error is a combination of strings of $X$ errors that start and end at the bottom boundary, strings of $Z$ errors that start and end at the left boundary, and a string of $X$ errors that starts at the bottom boundary and ends at the top boundary (\ie a \textbf{logical $\lx$ operator}).
	\end{itemize}
	
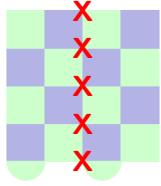
\begin{figure*}
		\centering\begin{tikzpicture}[x=0.5cm,y=0.5cm]
			\surfaceCodeCheckerboard{5}{5}{false}{false}{false}
			
			\foreach \x in {0,2} {
				\filldraw[draw=xStabColor,fill=xStabColor]
				(\x, 0) arc (-180:0:0.5) -- cycle;
			}
			
			\foreach \y in {0,...,4} \node at (2, \y) [errorStyle] {X};
			
		\end{tikzpicture}
		\caption{The change in the error for an input variable gadget that results when we change the value of that input variable.}\label{variableGadgetChange}
	\end{figure*}

	We analyze the change in the error at each gadget when we change the value of an input variable. We start with the input variable gadget. The change in the error when we change the value of that input variable is given in \figref{variableGadgetChange}; it is a string of $X$ errors that starts at the bottom boundary and continues upwards into the rest of the circuit.
	
\begin{figure*}
		\centering\begin{tikzpicture}[x=0.5cm,y=0.5cm]
			\surfaceCodeCheckerboard{19}{9}{false}{false}{false}
			
			\foreach \y in {0,1,2,4,6,7} \node at (7, \y) [errorStyle] {X};
			\foreach \x in {8,9,10} \node at (\x, 7) [errorStyle] {X};
			\foreach \y in {6,4,2} \node at (11, \y) [errorStyle] {X};
			
			\node at (3,3) [errorStyle] {Y};
			\node at (7,3) [errorStyle] {Y};
			\node at (11,3) [errorStyle] {Y};
			\node at (15,3) [errorStyle] {Y};
			\node at (3,5) [errorStyle] {Y};
			\node at (7,5) [errorStyle] {Y};
			\node at (11,5) [errorStyle] {Y};
			\node at (15,5) [errorStyle] {Y};
			
			\foreach \y in {1,2,4,6,7,8} \node at (3, \y) [errorStyle] {X};
			
			\foreach \y in {1,2,4,6,7,8} \node at (15, \y) [errorStyle] {X};
			
			\foreach \x in {1,2,4,5,6,8,9,10,12,13,14,16} \node at (\x, 3) [errorStyle] {Z};
			\node at (17,4) [errorStyle] {Z};
			\foreach \x in {17,16,14,13,12,10,9,8,6,5,4,2,1} \node at (\x, 5) [errorStyle] {Z};
			
			
			\node at (-2.2, 3) [labelStyle] {to left boundary $\longleftarrow$};
			\node at (-2.2, 5) [labelStyle] {to left boundary $\longleftarrow$};
			\node at (3, 0) [labelStyle] {$\downarrow$ \\ to bottom boundary};
			\node at (11, 1) [labelStyle] {$\downarrow$ \\ to bottom boundary};
			\node at (15, 0) [labelStyle] {$\downarrow$ \\ to bottom boundary};
			\node at (7,-1) [labelStyle] {$\uparrow$ \\ input wire};
			\node at (3,8.5) [labelStyle] {output 1};
			\node at (15,8.5) [labelStyle] {output 2};
		\end{tikzpicture}
		\caption{The change in the error for a FAN-OUT gadget that results when its input is changed.}\label{fanoutGadgetChange}
	\end{figure*}
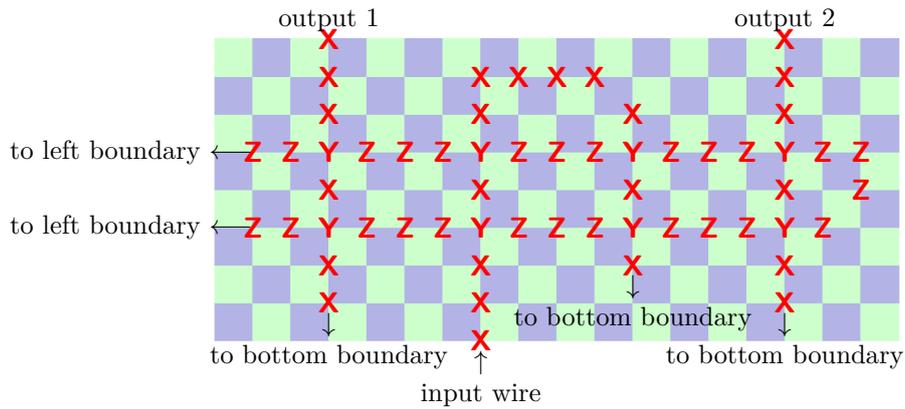

	When that string of $X$ error changes propagates up into a FAN-OUT gadget, the resulting change in the error for that gadget is in \figref{fanoutGadgetChange}. This change in the error is a string of $X$ errors that starts at the bottom of the gadget (\ie comes from the bottom boundary) and goes to the bottom boundary, and a string of $Z$ errors that starts and ends at the left boundary, and 2 strings of $X$ errors that start at the bottom boundary and continue upwards to the rest of the circuit.
	
	The NOT gadget is similarly trivial. When its input is changed, the change in the error for that gadget is a string of $X$ errors that starts at the bottom of that gadget (\ie comes from the bottom boundary) and continues upwards to the rest of the circuit. The AND gadget, however, has more complicated behavior when one or both of its inputs is changed, and it requires case analysis. We consider the case when its left input variable is changed; essentially the same argument applies if its right input variable is changed. This case breaks down into 2 sub-cases---the case when the right input variable is true, and the case when the right input variable is false. If both inputs are changed, the resulting change in the error for this gadget is the combination of error changes that would happen if the left input were changed then the right input were changed.
	\subsection{Case 1: left input variables is changed, right input variable is false}
		\begin{figure}
			\centering\begin{tikzpicture}[x=0.45cm,y=0.45cm]
				\andGadget
				
				\node at (12,20) [errorStyle] {Z};
				\node at (11,21) [errorStyle] {Z};
				\node at (10,22) [errorStyle] {Z};
				\node at (9,23) [errorStyle] {Z};
				\foreach \x in {8,7,6} \node at (\x,23) [errorStyle] {Z};
				
				\foreach \y in {0,...,4}  \node at (3, \y) [errorStyle] {X};
				\node at (3, 5) [errorStyle] {Y};
				\node at (3, 6) [errorStyle] {X};
				\node at (3, 7) [errorStyle] {Y};
				\foreach \y in {8,9} \node at (3, \y) [errorStyle] {X};
				
				\foreach \x in {4,...,9} \node at (\x, 9) [errorStyle] {X};
				\foreach \y in {10,11,12} \node at (10,\y) [errorStyle] {X};
				\node at (10,13) [errorStyle] {Y};
				\foreach \y in {14,15} \node at (10,\y) [errorStyle] {Z};
				\node at (10,16) [errorStyle] {Y};
				\node at (10,17) [errorStyle] {Z};
				\node at (11,18) [errorStyle] {Z};
				\node at (12,19) [errorStyle] {Z};
				
				\foreach \x in {9,8} \node at (\x,13) [errorStyle] {X};
				\foreach \y in {14,15} \node at (7,\y) [errorStyle] {X};
				\foreach \x in {8,9} \node at (\x,16) [errorStyle] {X};
				\node at (11,16) [errorStyle] {X};
				\node at (12,15) [errorStyle] {X};
				\foreach \y in {14,...,8} \node at (12,\y) [errorStyle] {X};
				\node at (12,6) [errorStyle] {X};
				\foreach \y in {4,...,1} \node at (12,\y) [errorStyle] {X};
				
				\foreach \x in {9,...,2} \node at (\x, 12) [errorStyle] {Z};
				\foreach \y in {11,...,7} \node at (1,\y) [errorStyle] {Z};
				\node at (2,7) [errorStyle] {Z};
				\foreach \x in {4,...,11} \node at (\x,7) [errorStyle] {Z};
				\node at (12,7) [errorStyle] {Y};
				\foreach \x in {13,...,25} \node at (\x,7) [errorStyle] {Z};
				\node at (26, 6) [errorStyle] {Z};
				\foreach \x in {26,...,13} \node at (\x,5) [errorStyle] {Z};
				\node at (12,5) [errorStyle] {Y};
				\foreach \x in {11,...,4} \node at (\x,5) [errorStyle] {Z};
				\foreach \x in {2,1} \node at (\x,5) [errorStyle] {Z};
			\end{tikzpicture}
			\caption{The change in the error for an AND gadget that results when its left input variable is changed and its right input variable is false.  This error displayed is the difference between the errors in \figref{andGadgetFF} and \figref{andGadgetTF}.}\label{andGadgetChangeFalse}
		\end{figure}
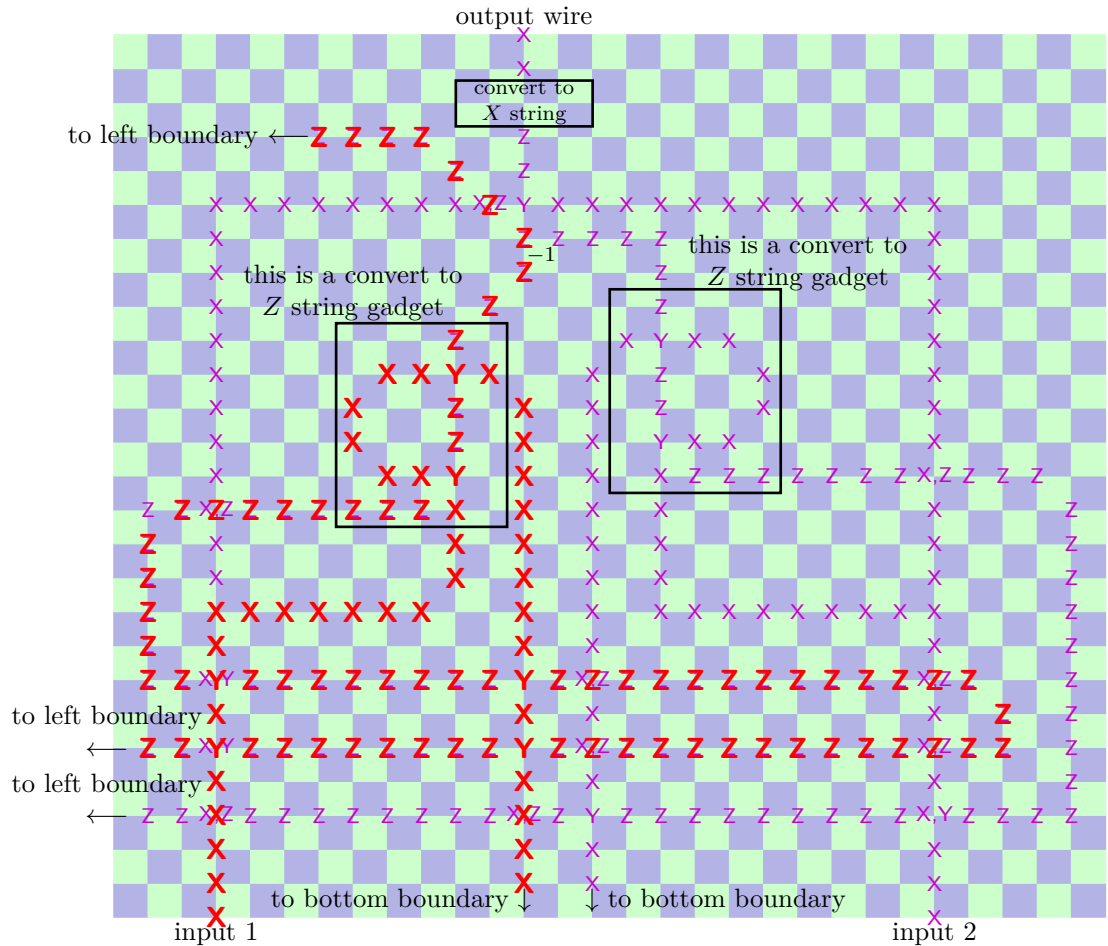
		\figref{andGadgetChangeFalse} shows the change in the error for an AND gadget when the left input's value is changed and the right input is false.  This is the operation that would have to be applied to change between the error in \figref{andGadgetFF} and \figref{andGadgetTF}, up to a global phase. This change in the error is a combination of an $X$ string that starts at the left input (\ie comes from the bottom boundary) and ends at the bottom boundary, and a $Z$ string that starts and ends at the left boundary. This change does not affect the output of this AND gadget.
	\subsection{Case 2: left input variables is changed, right input variable is true}
		\begin{figure}
			\centering\begin{tikzpicture}[x=0.45cm,y=0.45cm]
				\andGadget
				
				\node at (12,21) [errorStyle] {Y};
				
				\foreach \y in {0,1,2}  \node at (3, \y) [errorStyle] {X};
				\node at (3, 3) [errorStyle] {Y};
				\foreach \y in {4,...,21} \node at (3, \y) [errorStyle] {X};
				\foreach \x in {4,...,11} \node at (\x, 21) [errorStyle] {X};
				
				\foreach \y in {20,22,23} \node at (12,\y) [errorStyle] {Z};
				\foreach \y in {25,26} \node at (12,\y) [errorStyle] {X};
				
				\node [errorStyle] at (4, 24) {Z string to left boundary $\longleftarrow$};
				\node [errorStyle] at (20.5, 24) {$\searrow$ X string to bottom boundary};
				
				\foreach \y in {9,...,12} \node at (24, \y) [errorStyle] {X};
				\node at (24, 13) [errorStyle] {Y};
				\foreach \y in {14,...,21} \node at (24, \y) [errorStyle] {X};
				\foreach \x in {23,...,13} \node at (\x, 21) [errorStyle] {X};
				
				\foreach \x in {23,...,16} \node at (\x, 9) [errorStyle] {X};
				\foreach \y in {10,...,13} \node at (16, \y) [errorStyle] {X};
				\node at (16, 14) [errorStyle] {Y};
				\foreach \y in {15,16} \node at (16, \y) [errorStyle] {Z};
				\node at (16, 17) [errorStyle] {Y};
				\foreach \y in {18,19,20} \node at (16, \y) [errorStyle] {Z};
				\foreach \x in {15,14,13} \node at (\x, 20) [errorStyle] {Z};
				
				\foreach \x in {17,...,23} \node at (\x, 13) [errorStyle] {Z};
				\foreach \x in {25,26,27} \node at (\x, 13) [errorStyle] {Z};
				\foreach \y in {12,...,3} \node at (28, \y) [errorStyle] {Z};
				\foreach \x in {27,...,15} \node at (\x, 3) [errorStyle] {Z};
				\node at (14, 3) [errorStyle] {Y};
				\foreach \x in {13,...,4} \node at (\x, 3) [errorStyle] {Z};
				\foreach \x in {2,1} \node at (\x, 3) [errorStyle] {Z};
				
				\foreach \x in {17,18} \node at (\x, 14) [errorStyle] {X};
				\foreach \y in {15,16} \node at (19,\y) [errorStyle] {X};
				\foreach \x in {18,17} \node at (\x, 17) [errorStyle] {X};
				\node at (15, 17) [errorStyle] {X};
				\foreach \y in {16,...,4} \node at (14,\y) [errorStyle] {X};
				\foreach \y in {2,1} \node at (14,\y) [errorStyle] {X};
				
			\end{tikzpicture}
			\caption{The change in the error for an AND gadget that results when its left input variable is changed and its right input variable is true. This error displayed is the difference between the errors in \figref{andGadgetFT} and \figref{andGadgetTT}. The extra possible $X$ and $Z$ error strings coming out of the ``convert to $X$ string'' gadget (defined in \figref{gadgetsConvertXZ}) were implicit, and were not drawn on the original AND gadget (\figref{andGadget}) for clarity.}\label{andGadgetChangeTrue}
		\end{figure}
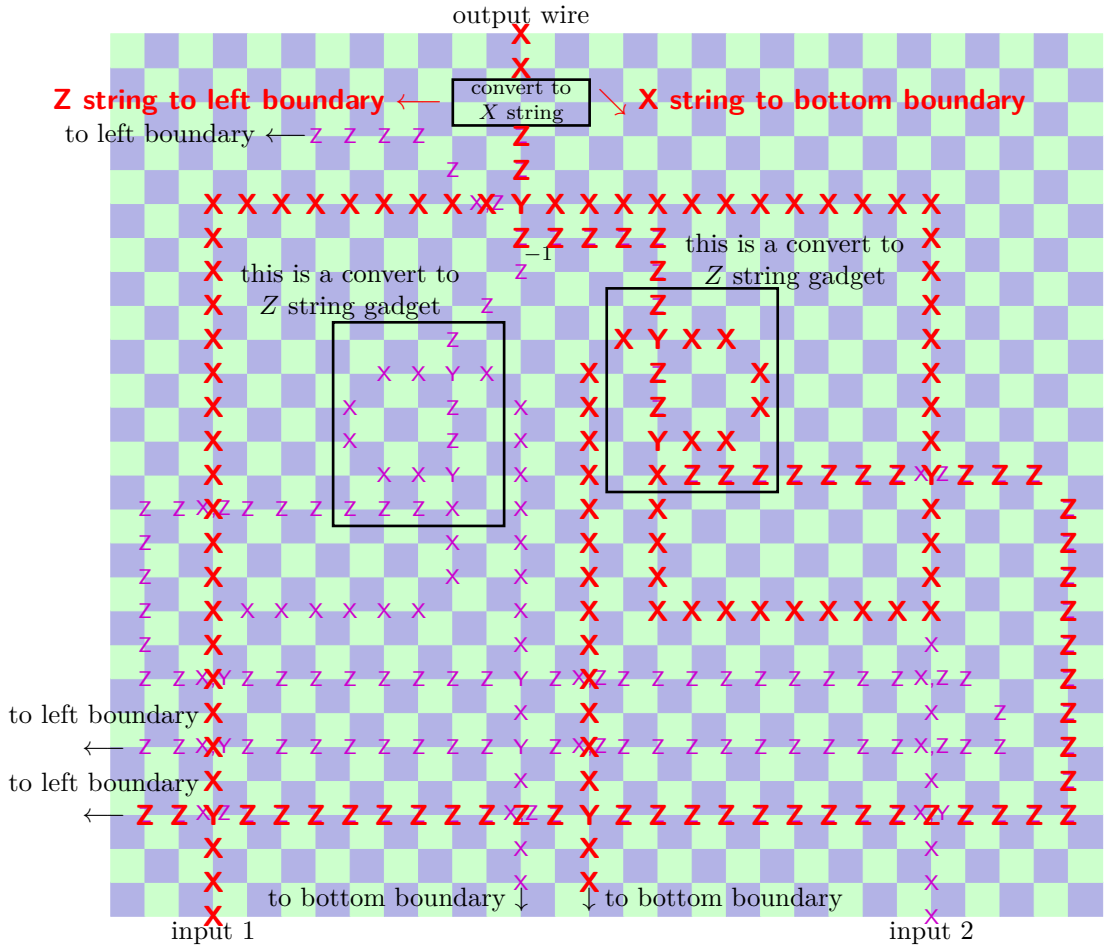
		\figref{andGadgetChangeTrue} shows the change in the error for an AND gadget when the left input's value is changed and the right input is true. This is the operation that would have to be applied to change between the error in \figref{andGadgetFT} and \figref{andGadgetTT}, up to a global phase. This change in the error is a combination of:
		\begin{itemize}
			\item An $X$ string that starts at the left input (\ie comes from the bottom boundary) and ends at the bottom boundary.
			\item A $Z$ string that starts and ends at the left boundary.
			\item An $X$ string that starts at the bottom boundary, comes up to the ``convert to $X$ string'' gadget, and continues up the output wire to the rest of the circuit.
		\end{itemize}
		This change does affect the output of this AND gadget. Note that the extra possible $X$ and $Z$ error strings coming out of the ``convert to $Z$ string'' gadget were implicit, and were not drawn on the original AND gadget (\figref{andGadget}) for clarity.
	\subsection{Putting it all together: change in the error for the whole circuit}
		Essentially the same argument holds if the right input variable of the AND gadget is changed instead of the left: then, the change in the error is strings of $X$ and $Z$ errors starting and ending at the same boundary, plus an $X$ string starting at the bottom boundary and ending at the output of the gadget iff the output of the gadget is changed. If both input variables of the AND gadget are changed, we can equivalently think of that as changing the left input variable and then the right input variable. Then the same result holds: the difference between those 2 errors is strings of $X$ and $Z$ errors starting and ending at the same boundary, plus an $X$ string starting at the bottom boundary and ending at the output of the gadget iff the output of the gadget is changed.
		
		Now that we have seen how all the gadgets change their error when their inputs are changed, we can consider the change in the error for the entire circuit (ie, all gadgets) if we change the value of one variable. If this change \textbf{does not} change the value of the output wire, then the total change in the error is a combination of:
		\begin{itemize}
			\item Many strings of $X$ errors that start and end at the bottom boundary, \ie a product of $X$ stabilizers.
			\item Many strings of $Z$ errors that start and end at the left boundary, \ie a product of $Z$ stabilizers.
		\end{itemize}
		If this change \textbf{does} change the value of the output wire, then the total change in the error is a combination of:
		\begin{itemize}
			\item Many strings of $X$ errors that start and end at the bottom boundary, \ie a product of $X$ stabilizers.
			\item Many strings of $Z$ errors that start and end at the left boundary, \ie a product of $Z$ stabilizers.
			\item A string of $X$ errors that starts at the bottom boundary and ends at the top boundary, \ie a \textbf{logical $\lx$ operator}.
		\end{itemize}
		This proves Lemma \ref{dqmldLemma}.
\section{Equivalent results for the standard, unrotated surface code}\label{unrotatedAppendix}
	In this paper we have worked with, and stated all of our results about, the rotated variant of the surface code\cite{bombin2007optimal}. Here we show that the same results hold in the standard, unrotated surface code.
	
	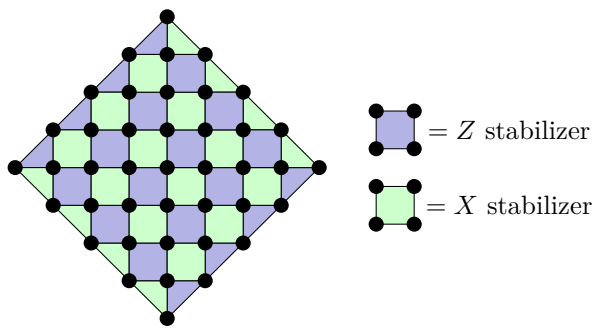
\begin{figure}
		\centering\begin{tikzpicture}[x=0.5cm,y=0.5cm]
		
			\foreach \x in {0,...,3} {
				\filldraw[draw=black,fill=zStabColor] (-4+\x,\x) -- (-3+\x,\x) -- (-3+\x,\x+1) -- cycle;
				\filldraw[draw=black,fill=zStabColor] (\x,-4+\x) -- (\x,-3+\x) -- (1+\x,-3+\x) -- cycle;
				
				\filldraw[draw=black,fill=xStabColor] (-4+\x,-\x) -- (-3+\x,-\x) -- (-3+\x,-\x-1) -- cycle;
				\filldraw[draw=black,fill=xStabColor] (\x,4-\x) -- (\x,3-\x) -- (1+\x,3-\x) -- cycle;
			}
			
			\foreach \x in {-3,...,2} {
				\pgfmathsetmacro{\startLoop}{\x < 0 ? -3 + (-\x-1) : -3 + \x};
				\pgfmathsetmacro{\endLoop}{\x < 0 ? 2 - (-\x-1) : 2 - \x};
				\foreach \y in {\startLoop,...,\endLoop} {
					\pgfmathsetmacro\stabColor{iseven(\x+\y) ? "zStabColor" : "xStabColor"}
					\draw[fill=\stabColor, draw=black] (\x, \y) rectangle ++(1,1);
					\node at (\x,\y) [qubitStyle] {};
				}
			}
			
			\foreach \x in {0,...,4} {
				\node at (-4+\x, \x) [qubitStyle] {};
				\node at (\x, -4+\x) [qubitStyle] {};
			}
			
			\foreach \x in {0,...,3} {
				\node at (\x, 3-\x) [qubitStyle] {};
			}
			
			\foreach \x in {1,...,3} {
				\node at (\x, 4-\x) [qubitStyle] {};
			}
		
			\draw[fill=xStabColor, draw=black] (5.5, -1.5) rectangle ++(1,1);
			\foreach \x in {5.5, 6.5} \foreach \y in {-1.5, -.5} {
				\node at (\x,\y) [qubitStyle] {};
			}
			\node at (9, -1) [labelStyle] {$= X$ stabilizer};

			\draw[fill=zStabColor, draw=black] (5.5, .5) rectangle ++(1,1);
			\foreach \x in {5.5, 6.5} \foreach \y in {.5, 1.5} {
				\node at (\x,\y) [qubitStyle] {};
			}
			\node at (9, 1) [labelStyle] {$= Z$ stabilizer};
		\end{tikzpicture}
		\caption{The $5\times 5$ unrotated surface code. The rotated and unrotated surface code have boundaries whose angles differ by $45\degree$. The boundaries of the unrotated surface code have 3-qubit $X$ and $Z$ stabilizers, instead of 2-qubit stabilizers. Compare with \figref{surfaceCodeDef}. This formulation of the unrotated surface code is equivalent to the standard formulation involving qubits at the edges of a square lattice, $Z$-stabilizers at faces of the lattice, $X$-stabilizers at points of the lattice, and smooth and rough boundaries\cite{surfaceCodeBoundaryBravyiKitaev}.}\label{unrotatedSurfaceCode}
	\end{figure}

	The unrotated surface code differs in the location and composition of its boundaries. The rotated and unrotated surface codes have boundaries whose angles differ by $45\degree$. Additionally, the unrotated surface code's boundaries consist of 3-qubit $Z$ and $X$ stabilizers (at what are often called the smooth and rough boundaries, respectively), instead of 2-qubit $Z$ and $X$ stabilizers. See \figref{unrotatedSurfaceCode} and compare with \figref{surfaceCodeDef}.
	
	The rotated surface code is generally preferred in practice because its rate $\frac{k}{n}$ is higher than that of the unrotated surface code, by a factor of 2.
	
	We can embed the same noise model defined by our gadgets into the unrotated surface code; the only subtlety is that we need to route all of the $X$ and $Z$ strings to an appropriate boundary where they can run to the boundary without causing a $-1$ syndrome measurement at the endpoint of the string. The unrotated surface code still has these boundaries; see \figref{unrotatedStringRouting} to see how we can embed our noise models into the unrotated surface code and still route the $X$ and $Z$ strings to appropriate boundaries. This only increases the size of the reduction by a constant factor.
	
	\begin{figure}
		\centering\begin{tikzpicture}[x=0.5cm,y=0.5cm]
			\draw (10, 0) -- (0, 10) -- (-10, 0) -- (0, -10) -- (10, 0);
			
			\node [rotate=-45] at (5.5, 5.5) {boundary with 3-qubit $X$ stabilizers};
			\node [rotate=45] at (-5.5, 5.5) {boundary with 3-qubit $Z$ stabilizers};
			\node [rotate=-45] at (-5.5, -5.5) {boundary with 3-qubit $X$ stabilizers};
			\node [rotate=45] at (5.5, -5.5) {boundary with 3-qubit $Z$ stabilizers};
			
			\node [align=center] at (-3, 2) {original \\ reduction};
			\draw (-5, 0) rectangle ++(4,4);
			
			\foreach \x in {-4.8,-4.6,...,-1.2} {
				\draw (\x, 0) -- (\x, -10-\x);
			}
			
			\foreach \y in {0.2,0.4,...,3.8} {
				\draw (-5, \y) -- (-10+\y, \y);
			}
		
			\draw (-3, 4) -- (-3, 6) -- (2, 6) -- (2, 8);
			
			\node at (0.2, -12) {$X$ strings ending at bottom of original reduction};
			\draw[->] (0.4, -11) -- (0.4, -7) -- (-1, -7);
			
			\node[align=left] at (-15, -0.4) {$Z$ strings ending at \\ left of original reduction};
			\draw[->] (-12, -0.4) -- (-7, -0.4) -- (-7, 0);
			
			\node at (0.4, 12) {output wire of original reduction};
			\draw[->] (0.4, 11.5) -- (0.4, 6.2);
		\end{tikzpicture}
		\caption{All of the $X$ and $Z$ strings that are routed to a boundary in the original reduction can be routed to an appropriate boundary in the unrotated surface code. The $X$ strings that end at the bottom boundary of the rotated surface code will end at the bottom-left boundary of the unrotated surface code. The $Z$ strings that end at the left boundary of the rotated surface code will end at the top-left boundary of the unrotated surface code. The output wire of the original reduction, which ends at the top boundary of the rotated surface code, will end at the top-right boundary of the unrotated surface code.}\label{unrotatedStringRouting}
	\end{figure}

\section{Completeness results}
	\label{completenessAppendix}

	We would like to show that QMLD and DQMLD are not only hard, but complete for NP and \#P, respectively. However, QMLD and DQMLD as we have formulated them are are function problems, not decision problems, as the correct answers to the problems are Pauli strings rather than a yes/no answer. Giving completeness results for a function problem requires 4 steps:
	\begin{enumerate}
		\item Reformulating the function problem as a decision problem.\label{reformulateStep}
		\item Showing that the decision problem is in the relevant complexity class (in this case, NP or $\mathrm{P}^{\mathrm{\#P}}$).\label{inClass}
		\item Showing that the decision problem maintains the same hardness results as the function problem.\label{maintainsHard}
		\item Showing that the function and decision variants of the problem are equivalent, in the sense that an oracle for the decision problem can be used to solve the function problem in polynomial time.\label{functionDecisionEquivalent}
	\end{enumerate}
	
	Technically, DQMLD (reformulated as a decision problem) becomes complete for $\mathrm{P}^{\mathrm{\#P}}$ (the set of decision problems solvable in polynomial time with an oracle for \#P), not \#P. This is because the problem can't be formulated as a function problem where the output of the function is the number of accepting paths for some nondeterministic TM, which is required to place a problem in \#P.
	
	\subsection{Completeness result for QMLD}
		The decision problem variant of QMLD is: given a surface code instance, a noise model, a set of syndromes, and a probability $p'$, does there exist a Pauli error consistent with the syndromes with probability at least $p'$? We require the error probabilities in the noise model, and $p'$, to be rational numbers. This reformulation as a decision problem is step \ref{reformulateStep}. This problem is in NP, because one can give a Pauli error with a high enough probability as a witness, which can easily be verified. This completes step \ref{inClass}.
		
		A reduction from SAT to this decision problem uses the same surface code instance, noise model, and syndromes as the original QMLD reduction, and also outputs $p'=p^{\ell}$. Recall that all errors corresponding to satisfying assignments have probability $>p^{\ell}$ (see \eqref{riggedProbabilities}), and all errors corresponding to unsatisfying assignments have probability $<p^{\ell}$. If the answer to the decision problem with $p'=p^{\ell}$ is yes, then such an error with probability $>p'=p^{\ell}$ corresponds to a satisfying assignment and the formula is satisfiable; if the answer to that decision problem is no, then all errors have probability $<p'=p^{\ell}$ and thus the formula is unsatisfiable. Thus this decision problem is NP-hard; this completes step \ref{maintainsHard}. Note that this is a many-one/Karp reduction, in contrast to the weaker Turing reductions used in the main text.
		
		\subsubsection{Equivalence between the decision problem and the function problem}
			We give a polynomial-time algorithm that uses an oracle for the decision problem to solve the function problem (\ie actually finding a maximum probability error\footnote{We say ``a maximum probability error'' rather than ``the maximum probability error'' because there may be multiple errors that are tied for the highest probability. The task of the function problem is to just output one of them.}). This algorithm has 2 steps:
			\begin{enumerate}
				\item Determine the probability of a maximum probability error. \label{findProbStep}
				\item Find an error with such maximum probability. \label{findErrorStep}
			\end{enumerate}
			
			Let the error probabilities in the noise model be rational numbers $a_i/b_i$, where $a_i$ and $b_i$ are integers and $i$ is an index for a qubit and a Pauli $X$, $Y$, $Z$. The probability of any Pauli error is some integer divided by $\prod_i b_i$. This integer $\prod_i b_i$, which we call $b$, can be calculated in time that is polynomial in the size of the noise model. Let the probability of a maximum probability error be $a/b$. We can determine $a$, bit-by-bit, using binary search, with $O(\log b)$ calls to the oracle. This is step \ref{findProbStep} of the algorithm.
			
			Once we know the probability of a maximum probability error, we can find an error with that probability qubit-by-qubit in the following way. Let the probability of a maximum probability error be $P$. We start with qubit 0: for each Pauli $k\in\{\mathbb{I},X,Y,Z\}$ with nonzero error probability on qubit 0, we can query the decision problem oracle with a modified noise model that sets $p^{(0)}_k=1$ (\ie makes it so the Pauli error $k$ occurs on qubit 0 with probability 1) and with a new $p'$ equal to $P/p^{(0)}_k$ (here $p^{(0)}_k$ is the original $p^{(0)}_k$ before we set it to 1). Whichever of those oracle calls returns yes tells us whether there is an error with probability $P$ with the Pauli error $k$ occurring on qubit 0. Once we have found such a Pauli error $k$ that occurs on qubit 0 in a maximum probability error, we modify our noise model to set $p^{(0)}_k=1$ and note that the probability of a maximum probability error in this new noise model is $P/p^{(0)}_k$ (again, here $P$ and $p^{(0)}_k$ are the originals before modification). We then use this new noise model and new probability of a maximum probability error to determine the Pauli error for qubits 1, 2, etc., until we have determined the entire error. This completes step \ref{functionDecisionEquivalent}.
		
	\subsection{Completeness result for DQMLD}
		The decision problem variant of DQMLD is: given a surface code instance with stabilizer group $\mathcal S$, a noise model, a set of syndromes, and an error $E$ consistent with the syndromes, is a maximum likelihood coset $E\mathcal S$\footnote{Again, we say ``a maximum likelihood coset'', rather than ``the maximum likelihood coset'', because there may be multiple cosets tied for the highest coset probability.}? This reformulation as a decision problem is step \ref{reformulateStep}.
		
		We can use an oracle for this decision problem to solve the \#P-complete problem \#SAT (and thus, any problem in $\mathrm{P}^{\mathrm{\#P}}$) in polynomial time using the exact same strategy as in the proof of \#P-hardness of the function problem in the main text. We only need to supply to the oracle an additional Pauli error $E$ made from some arbitrary assignment of the variables of the formula. So this decision problem maintains the same hardness result (\#P-hard, and also hard for $\mathrm{P}^{\mathrm{\#P}}$, under Turing reductions), which completes step \ref{maintainsHard}.
		
		Technically, this decision problem cannot be in \#P itself, because \#P is the set of function problems whose output is the number of accepting paths of a nondeterministic Turing machine, and it does not appear to be possible to formulate the DQMLD problem in a way such that the right answer takes that form. So instead we put this decision problem in the more-or-less equivalent complexity class $\mathrm{P}^{\mathrm{\#P}}$.
		
		\subsubsection{Solving DQMLD with a \#P oracle}
			We give a polynomial-time algorithm that uses an oracle for \#P to solve the DQMLD decision problem. We start with any particular error $E$ consistent with the syndromes, which can be found in polynomial time. Consider a randomized algorithm that samples Pauli errors from the noise model, throws out errors inconsistent with the syndromes (by accepting or rejecting with probability $\frac{1}{2}$), and accepts if the error is equivalent to $E$ up to stabilizers (which can be determined in polynomial time). We can use a \#P oracle to determine the acceptance probability of this algorithm. We determine the acceptance probability of that randomized algorithm with $E$, $\bar X E$, $\bar Z E$, and $\bar X\bar Z E$. The maximum of those acceptance probabilities tells us a maximum likelihood coset. This puts DQMLD in $\mathrm{P}^{\mathrm{\#P}}$ and completes step \ref{inClass}.
			
		\subsubsection{Equivalence between the decision problem and the function problem}
			The function problem is simply outputting an error in a maximum likelihood error coset. Given any error $E$ consistent with the syndromes (which can be found in polynomial time), we can use the oracle to determine whether $E$, $\bar X E$, $\bar Z E$, or $\bar X\bar Z E$ is in a maximum likelihood coset (at least one of those must be the case). We then output whichever one of those errors is in a maximum likelihood coset. This completes step \ref{functionDecisionEquivalent}.

\end{appendices}

\end{document}